%% file: ArinSym.tex
\documentclass[notitlepage,16pt]{article}
\usepackage[utf8]{inputenc}\usepackage{pdfsync}\usepackage{float}
\usepackage[T1]{fontenc}
\usepackage{amsfonts,amsmath,amssymb,mathrsfs,url,authblk,theorem,
geometry,esint,subcaption,
comment
}
\usepackage{bbm}
\usepackage{array}
\usepackage{tabularx}
\usepackage[utf8]{inputenc}
\usepackage{todonotes}

\usepackage{graphicx}
\graphicspath{{figures/}}
\DeclareGraphicsExtensions{.pdf,.png}
\usepackage[normalem]{ulem}
\usepackage{hyperref}
\usepackage{cleveref}

\usepackage{color}
\usepackage{textcomp}
\usepackage{geometry}
\usepackage{enumitem}

\usepackage{epsfig}
\usepackage{wasysym}
\usepackage{empheq}
\usepackage{url}
\usepackage{tikz}
\usetikzlibrary{positioning}
\usetikzlibrary{arrows}
\usetikzlibrary{arrows.meta}
\usetikzlibrary{calc}
\newtheorem{theorem}{Theorem}
\newtheorem{remark}{Remark}
\newtheorem{proposition}{Proposition}
\newtheorem{definition}{Definition}
\newtheorem{corollary}{Corollary}
\newtheorem{example}{Example}
\newtheorem{lemma}{Lemma}
\newtheorem{assumption}{Assumption}

 \def\eeD{\end{definition}} \def\beD{\begin{definition}}
\def\beR{\begin{remark}} \def\eeR{\end{remark}}
\def\beL{\begin{lemma}} \def\eeL{\end{lemma}}
\def\beC{\begin{corollary}
  }\def\eeC{\end{corollary}}
  \def\beT{\begin{theorem}}\def\eeT{\end{theorem}}
  \def\beP{\begin{proposition}} \def\eeP{\end{proposition}}
\def\beXa{\begin{example}} \def\eeXa{\end{example}}
\def\beA{\begin{assumption}} \def\eeA{\end{assumption}}

\newcommand{\RN}[1]{\textup{\uppercase\expandafter{\romannumeral#1}}}

\def\mF{\mathcal F} 

\newcommand{\al}{\alpha}

\renewcommand{\L}{\Lambda}
\newcommand{\de}{\partial}

\newcommand{\R}{\mathbb{R}}
\newcommand{\N}{\mathbb{N}}

\def\How{However, }
\def\no{\nonumber}\def\bep{\begin{pmatrix}} \def\eep{\end{pmatrix}}
\def\bev{\begin{vmatrix}} \def\eev{\end{vmatrix}}

\def\l{{\lambda}}\def\Lgg{\Lambda +\gamma _r+\gamma _s}
\def\PF{Perron Frobenius } 

\providecommand{\pp}[1]{\left[#1\right]} 
\providecommand{\pr}[1]{\left(#1\right)} 

\def\QED{\hfill {$\square$}\goodbreak \medskip}

\renewcommand{\theta}{\vartheta}

\renewcommand{\thefootnote}{\fnsymbol{footnote}}

\numberwithin{equation}{section}

\def\eig{eigenvalue}\def\satd{satisfied}\def\mD{{\mathcal D}} \def\mV{{\mathcal V}}

\def\ngm{next generation matrix}
\def\bc{\begin{cases}
  }     
\def\ec{\end{cases}}  
   \def\for{\forall}

  \newcommand{\beq}{\begin{eqnarray}
    }
\def\eeq{\end{eqnarray}}
   \newcommand{\be}[1]{\begin{equation}\label{#1}}
\newcommand{\ee}{\end{equation}}
\def\bea{\begin{eqnarray*}}\def\ssec{\subsection}
  \def\prf{{\bf Proof:} } \def\Prf{{\bf Proof:} } \def\satg{satisfying } 
\def\eea{\end{eqnarray*}} \def\la{\label}\def\fe{for example } \def\wlo{w.l.o.g.}   
   \def\Mp{More precisely, } \def\sats{satisfies}  \def\saty{satisfy}        

\def\I{\infty} \def\Eq{\Leftrightarrow}
  
\def\PH{phase-type }  
\def\BEN{\begin{enumerate}}  \def\BI{\begin{itemize}}
\def\EEN{\end{enumerate}}   \def\EI{\end{itemize}} \def\im{\item} \def\Lra{\Longrightarrow}  \def\eqr{\eqref}  
\def\no{\nonumber} 
\def\mR{\mathcal R} 
\def\g{\gamma}   \def\de{\delta}
   
\def\ep{\epsilon} 
   \def\al{\alpha}
\def\Fr{Furthermore, }
    
   \def\Mp{More precisely, } 
   \def\wrt{with respect to }
  \def\mbw{may be written as }\def\resp{respectively} \def\wlo{w.l.o.g. } \def\fno{from now on} 
 \def\eqr{\eqref}  
\def\wk{well-known}
\def\a{\alpha}
\def\l{\; \mathsf l}
\def\fr{\frac} \def\im{\item}
\newcommand{\s}{\;\mathsf s}
\renewcommand{\i}{\;\mathsf i}
\renewcommand{\r}{\; \mathsf r}

\def\eeD{\end{definition}} \def\beD{\begin{definition}}
\def\beR{\begin{remark}} \def\eeR{\end{remark}}
\def\beL{\begin{lemma}} \def\eeL{\end{lemma}}

        \def\ba{\bff \a} \def\vi{\vec \i} \def\bi{\bff \i} 
    \def\va{\vec a } 
     \newcommand{\bff}[1]{{\mbox{\boldmath$#1$}}}
    \def\beP{\begin{proposition}} \def\eeP{\end{proposition}}

    \long\def\symbolfootnote[#1]#2{
\begingroup
\def\thefootnote{\fnsymbol{footnote}}\footnote[#1]{#2}
\endgroup}
\def\fn{\symbolfootnote}
\newcommand{\e}{\;\mathsf e}
\def\corr{corresponding }
\def\para{parameter} \def\paras{parameters}
\def\l{\lambda} \def\si{\sigma}
 \def\vb{\vec \b}
 \def\FA{\bf FA} \def\IA{\bf IA}
\newcommand{\rt}{\right}\def\n{\de}
\def\D{\Delta} \def\elt{element}
\def\Itf{It follows that }
\newcommand{\red}{\textcolor[rgb]{1.00,0.00,0.00}}

 \def\com{compartment}\def\wmc{we may check that }
 
\newcommand{\realpos}{\mathbb{R}_{> 0}}
\newcommand{\realnneg}{\mathbb{R}_{\geq 0}}
\def\Lgn{\L+\g+\delta}\def\Lg{\L+\g_e}
\newcommand{\figu}[3]{
\begin{figure}[H]
\centering
\includegraphics[scale=#3]{#1}
\caption{#2\label{f:#1}}
\end{figure}
}

\def\mR{\mathcal R}\def\n{\mathsf n}\def\1{\mathbf{1}} \def\m0{{\mathcal R}_0}

\def\brn{basic reproduction number }
 \def\DFE{disease free equilibrium}\def\vn{\vec \de}  \def\vi{\vec \i}  \def\vz{\vec 0}\def\ie{\i_{ee}}\def\re{\r_{ee}}\def\se{\s_{ee}}\def\eE{\e_{ee}}\def\rd{\r_{dfe}} \def\sd{\s_{dfe}}
\def\b{\beta}  \def\ei{e_i} \def\SM{{\bf SM }} \def\FA{{\bf FA }} \def\IA{{\bf IA }}

\title{ New results and open questions for SIR-PH epidemic models with linear birth rate, loss of immunity, vaccination,  and   disease and vaccination fatalities}
\date{\today}

\begin{document}

 \author[1]{Florin Avram \footnote{Corresponding author. E-mail:
     Florin.Avram@univ-Pau.fr}}
 \author[2]{Rim Adenane}
 \author[3]{Andrei Halanay}
 \affil[1]{Laboratoire de Math\'{e}matiques Appliqu\'{e}es, Universit\'{e} de Pau, France 64000 }
 \affil[2]{D\'epartement des Math\'ematiques, Universit\'e Ibn-Tofail, Kenitra, Maroc 14000}
 \affil[3]{Department of Mathematics and Informatics,  Polytechnic University of Bucharest, 062203 Bucharest}
 \maketitle

\begin{abstract}
\section*{Abstract}
Our paper presents three new classes of models: SIR-PH, SIR-PH-FA, and
SIR-PH-IA, and states two problems we would like to solve about them. Recall
that deterministic mathematical epidemiology has one basic general law, the ``$\m0$
alternative" of \cite{Van, Van08}, which states that the local stability condition of the disease
free equilibrium may be expressed as $\m0<1$, where $\m0$ is the famous basic
reproduction number, which plays also a major role in the theory of branching
processes. The literature suggests that it is impossible to find general laws
concerning the endemic points. However, it is quite common that
\BEN
\im   When $\m0>1$, there exists a unique fixed endemic point, and 
\im  the endemic point is  locally stable when $\m0>1$.\EEN
One would like to establish these properties for a large class of realistic epidemic models (and we do not include here epidemics without casualties).
We have introduced in \cite{AAK, AABBGH} a ``simple", but broad class of ``SIR-PH models" with varying population, with the express purpose of establishing for these processes the two properties above. Since that seemed still hard, we have introduced a further class of ``SIR-PH-FA" models, which may be interpreted as approximations for the SIR-PH models, and which includes simpler  models typically studied in the literature (with constant population, without loss of immunity, etc). For this
class, the first ``endemic law" above is ``almost established", since explicit formulas for a unique endemic point are available, independently of the number of infectious compartments –see Proposition \ref{p:ee}, and it only remains to check its belonging to the invariant domain.  
This may yet turn out to be always verified, but we have not been able to establish that.
However, the second property, the sufficiency of $\m0>1$ for the local stability of an
endemic point, remains  open even for SIR-PH-FA models, despite the
numerous particular cases in which it was checked to hold (via Routh-Hurwitz
time-onerous computations, or Lyapunov functions). The goal of our paper is
to draw attention to the two open problems above, for the SIR-PH, SIR-PH-FA,
and also for a second, more refined ``intermediate approximation" SIR-PH-IA. We
illustrate the current status-quo by presenting new results on  a generalization of the SAIRS epidemic model
of \cite{RobSti, Ott}.

\end{abstract}

\textbf{keywords}:epidemic models; varying population models; SIR-PH models; stability; next-generation matrix approach; basic reproduction number; vaccination; loss of immunity; endemic equilibria; Routh-Hurwitz conditions.

\section{Introduction}

{\bf Motivation}. One of the hardest challenges facing epidemic  models is dealing
 with  models with in which death is possible,  in which  the total population $N$  varies, and in which the infection rates  depend on $N$ (as is the case in reality, except for a short period of time at the start of an epidemic).
Since these  features confront the researcher with   challenging behaviors (the first being that the  uniqueness of the fixed endemic point may stop holding), sometimes hard to explain epidemiologically,  it seems natural  to attempt to identify the simplest class of realistic  models for which  a theory may be developed. The natural choice is ``standard  incidence rates" -- see \eqr{SIRNe}, since models  with nonlinear infection rates are quite complex -- see \fe\ \cite{Liu1986,Liu87,Vyska,Gha,Gupta}  for the very complex dynamical behaviors which may arise otherwise.

The next issue is choosing the type of birth function $b(N)$ to work with. {The easiest case is when $b(N)$ is constant}, but this corresponds to immigration rather than birth, and so our favorite are linear birth rates $b(N)=\L N$. A bonus for this choice, as well known,
is that normalization by the total population leads  to a model with constant parameters, which looks similar to classic  constant population models,  but involves some extra nonlinear terms -- see \eqr{SEIRsc}. The well-studied classic models may then be recovered via a heuristic ``first  approximation" ({\bf FA}) of ignoring the extra terms.  This approximation, which deserves being  investigated  rigorously via slow-fast/singular perturbation/homogeneization techniques, has the merit of putting under one umbrella constant and varying population models with linear birth rates.

At this point, let us mention that we believe that epidemic models should be ideally parameterized by  the two matrices $F,V$ which intervene in the \ngm\ approach, which have been called  disease carrying and state evolution matrices \cite{De19, De21}. A foundational paper in this direction is \cite{Arino}, which show  that further simplifications arise   for models having only one susceptible class, and also disease carrying matrix of rank one. The first fundamental question, the uniqueness of the endemic point when $\m0>1$, may be resolved explicitly   for the  ``FA approximation" (and hence also  for ``small perturbations", numerically).
This has motivated us to propose in  \cite{AAK, AABBGH} to develop the theory of this class of models, which we  call ``Arino" or ``SIR-PH" models.

{\bf Contributions and contents}.
The goal of our paper is to draw attention to two interesting  open problems , for the SIR-PH, SIR-PH-FA,
and also for a second, more refined ``intermediate approximation" SIR-PH-IA. We
illustrate the current status-quo by presenting new results on  a generalization of the SAIRS epidemic model of \cite{RobSti, Ott}.

 The  SAIRS model \eqr{SEIRsc} is presented in Section \ref{s:SAIR}. The history of the problem and some oversights and errors in the literature are recalled in Section \ref{s:hys}.
{The basic reproduction number} $\m0$  and the weak $\m0$ alternative for the DFE equilibrium are established via the \ngm\ approach in Section \ref{s:R0}.
The local stability of the endemic point when the basic reproduction number satisfies $\m0>1$  for the \FA model  are established in Section \ref{s:LSFA}.

A review of the theory of SIR-PH models is provided in Section \ref{s:SIRPH}, and some new results in Section \ref{s:IA}.

The scaled SAIRS model   is revisited  in the Section \ref{s:ee} (Appendix), where some previous results in the literature are corrected and completed.
\input{SAIR1}

\input{SAIR2}

\input{SIRPH}

\input{SIR2}

\input{IA}

\input{SAIRsc}

\input{con1}

\section{Conclusions and further work}
Our paper highlighted  several open problems for SIR-PH, SIR-PH-IA and SIR-PH-FA  models. The following general directions seem worthy of further work.
\BEN \im Investigate whether the beautiful determinant formula \eqr{detO} hidden  in the papers of \cite{RobSti,Ott} (valid when $\de=0$) may be extended to SIR-PH models, exploiting the partition \eqr{Jac2}.
\im Study the case of two or more \com s susceptible to become infected (\fe\ the SEIT model \cite{Van08,Mart}).
\im Study the scaled  model as a perturbation of the \FA model.
\im Study   stability via the geometric approach of Li, Graef, Wang, Karsai, Muldoney and  Lu \cite{LiGraef}.
\im We hope that the use of more sophisticated and fast software will allow researchers in the future to progress with the interesting questions raised by models with higher dimensions. 
Here, exploiting symmetries may turn out helpful.


\EEN

{\bf Acknowledgement}. We thank Mattia Sensi and Sara Sottile for useful suggestions  and references, and for providing some of the codes  used for producing the figures.

\bibliographystyle{amsalpha}

\bibliography{Pare38}

\end{document}

%% file: SAIR1.tex
\section{The SAIRS model with linear birth rate \la{s:SAIR}}
{In this paper,} we consider a ten parameters SAIR (also called  SEIR in the classic literature \cite{Van08}) epidemic model inspired by \cite{Green97,LiGraef,LiMul0,SunHsieh,Brit18,LuLu,douris2019global,Ott}, which we call  SAIR/V+S (or SAIR for short), since it groups together immunized people in an R/V  \com. The letter A (from asymptomatic) stands for the fact that the individuals in this compartment may infect the susceptibles. This important feature, already present in  \cite{Van08}, was further studied in  \cite{RobSti,Ansumali,Ott}. The model {studied in \cite{Ott}} is the most complete in the sense that it misses only one of the parameters of interest, namely  the important
extra death rate due to the disease --see \fe\ \cite{Graef}. The explanation of this
omission  in \cite{Ott} lies probably in the fact that this paper follows the tradition of the ``short term constant population epidemics'', in which the total population $N$  is assumed constant and the endemic point is unique and easy to find explicitly. Our paper investigates, for the model of \cite{Ott}, the topic of the first six papers cited, i.e. we attempt to deal with the variation of $N(t)$ during  epidemics which may last for a long time, and which may never be totally eradicated.
We generalize the uniqueness of the endemic point and the local stability results of  these six papers, and at the same time draw attention to certain unnecessary assumptions  and mistakes.  The hardest issue, that of global stability, is only illustrated via some numerical simulations. 

{\bf SAIR Model}.
Letting  $S(t)$, $E(t)$, $I(t)$, $R(t)$, $D(t)$ and $D_e(t)$ represent Susceptible individuals, Exposed individuals, Infective individuals, Recovered individuals, naturally Dead individuals and Dead individuals due to the disease, respectively, the  model we consider is:
\begin{align}
\la{SIRNe}
S'(t)&=\L N(t) -S (t)\pr{\fr{\b_i}{N(t)}  I (t)+\fr{\beta_e}{N(t)}  E (t) +\g_s+ \mu} + \g_r R(t),
\nonumber\\
E'(t)&=S (t)\pr{\fr{\b_i}{N(t)}  I (t)+\fr{\beta_e}{N(t)}  E (t)}-(\g_e+\mu) E(t),\quad \g_e=\ei+e_r\nonumber\\
I'(t)&=\ei E(t)- I(t)\pr{\g  + \mu+\de},\nonumber\\
R'(t)&= 
\g_s S(t)+e_r E(t)+\g    I(t) -(\g_r+\mu) R(t), \nonumber\\
D'(t)&= \mu(S(t)+ E(t) + I(t)+R(t)):=\mu N(t),\nonumber\\
D_e'(t)&=\de  I(t),\nonumber \\
N(t)&=S(t)+E(t)+I(t)+R(t) \Lra N'(t)=(\L  -  \mu)  N(t) -\de  I(t).
\end{align}

{\bf Epidemiologic meaning of the parameters}:
\BEN \im  $\L \in \realnneg$  and $ \mu \in \realnneg$ denote the average
birth and death rates in the population (in the absence of
the disease), respectively.
\im {The parameters} $\b_i\in \realnneg$ and $\beta_e\in \realnneg$ denote the infection rates for infective and exposed individuals, \resp;
\im $\g_s \in \realnneg$  is  the  vaccination rate,
$\g_e \in \realnneg$  is  the  rate at which the exposed individuals become infected or recovered,
$\g_r \in \realnneg$  denotes the rate at which immune individuals
lose immunity (this is the reciprocal of the expected duration of
immunity),  $\g  \in \realpos$ is the  rate at which infected
individuals recover from the disease.
\im
 $\ei\in \realnneg$, $e_r\in \realnneg$, are rates of transfer from $E$ to $I$ and $R$, \resp.
 \im  $\de \in \realnneg$  is the extra death rate in  the infected \com\ due to the disease.
 \EEN

{\bf Some particular cases}.
\BEN \im  If $\b_e=e_r=0$, we obtain a SEIRS type model. If furthermore  $\g_r=\g_s=0$, we {arrive to} the model masterly
studied in \cite{LiGraef}; if only one of these parameters is zero, we {arrive to} the models
studied in \cite{SunHsieh} and \cite{LiMul0,Brit18,LuLu}.
\im If $\b_i=\g_r=\g_s=0=\de$, we obtain the  SIQR model \cite{SIQR,Russo}. \EEN

\beR \la{r:not} Note the notation scheme employed above, which could be applied to any compartmental model. A linear rate of transfer from compartment $m$ to compartment $c$ is denoted by $m_c$, and the total linear rate out of $m$ is denoted by $\g_m$,  which implies $\sum_{c} m_c= \g_m$, \fe\ $\g_e=\ei+e_r$. Extra death rates due to the epidemics
in a department $c$ are denoted by $\de_c$. An exception is made though for the infectious compartment $\i$, where we simply use the classic notations $\g,\de$, instead of $\g_i,\de_i$. Our scheme  would simplify a lot, if adopted,  perusing the rather random notations used in the literature. 
\eeR

{\bf The scaled model}. It is  convenient to reformulate \eqr{SIRNe} in terms of the
normalized fractions
$
\s=\fr{S}N,
\e=\fr{E}N,
\i=\fr{I}N,
\r=\fr{R}N.
$ Using $N'(t)=(\L-\mu)N(t)- \de I(t),$ this yields the following  nine parameters SEIRS epidemic model
(the common death rate $\mu$  simplifies,  and an extra $\de \i \; c$ appears in the equation of each compartment $c$--see \fe\cite{AABH} for similar computations).

\be{SEIRsc} \bc
\s'(t)= \L   -\s(t)\pr{\b_i  \i(t)+\beta_e \e(t)+\g_s+\L} + \g_r \r(t)+\de\s(t)  \i(t)\\
\bep \e'(t)\\\i'(t)\eep =
\pp{\s(t) \bep  \beta_e     &  {\b_i} \\
 0&  0 \eep+ \bep    -(\g_e +\L)  &  0 \\
 \ei&  -\pr{\Lgn}\eep+ \de \i_t Id}
  \bep \e(t)\\\i(t)\eep
\\
\r'(t)=  \g_s \s(t)+ e_r   \e(t)+ \g   \i(t)- (\g_r+\L) \r(t)+\de \i(t)\r(t) \ec.
\ee

\beR Note that we have written the ``infectious" middle equations to emphasize first the factorized form,  similar to that   encountered  for Lotka-Volterra networks -- see \fe\ \cite{goh1977global}.
Secondly, for the factor appearing in these equations, we have emphasized the form
 \be{V}F - V + \de \i I_2:=\s B - V + \de \i I_2.\ee

 Note that the matrices $F,V$ are featured in the famous the  {(Next Generation Matrix)} approach \cite{Van08}, that some authors refer to them as ``new infections" and ``transmission matrices",  that \cite{De21} call them the disease carrying and state evolution matrices, and  that \cite[Ch 5]{bacaer2021mathematiques} gives a way to define an associated stochastic birth and death model associated to these matrices.
 The matrix $B$  is further useful in defining and studying more general SIR-PH models -- see \cite{AAK,AABBGH}  and below,  and see also
   \cite{ballyk2005global}, \cite[(2.1)]{Fall}, \cite{De19}  for related works.

   Since the computations will become soon very cumbersome, we will start using \fno\ the following notations:
 \be{h}\bc v_0=\Lgg,\\ v_1=\Lg, \\ v_2=\Lgn. \ec\ee
 Note that $v_z, \mbox{with} \;z=\{1,2\},$ are the diagonal elements of $V$, and that $v_0$ appears as denominator in the DFE \eqr{sd}.

 \eeR

\begin{figure}[H]
\centering
\includegraphics[width=.8\columnwidth]{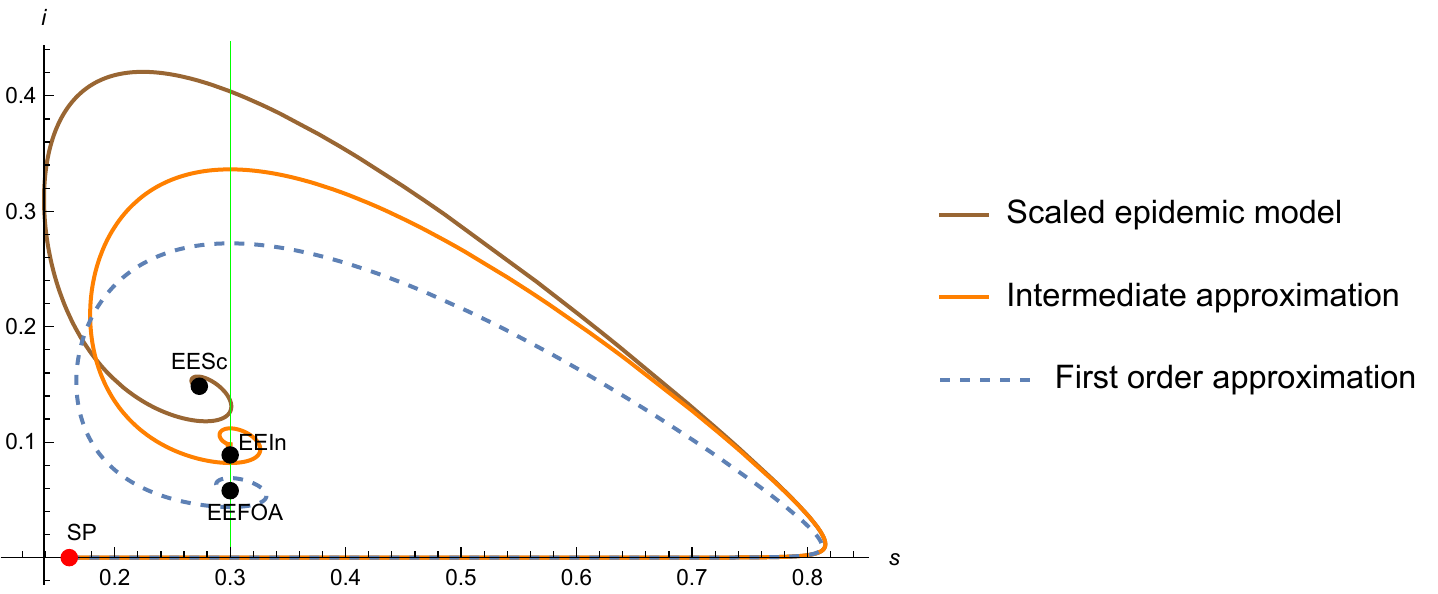}
 \caption{Parametric $(\s,\i)$ plots of the scaled epidemic and   its FA and intermediate approximations for a SIR-type model with one infectious class, starting from a starting point SP with $i_0=10^{-6}$, with
 $\m0=3.21, \b=5,\; \g  = 1/2,\; \L=\mu=1/10, \; \g_r=1/6, \; \g_s=.01, \;\de_i=.9, \; \de_r=0$. $EESc, EEIn,EEFOA$ are the stable endemic points of the scaled model, intermediate model, and the FA model, respectively.
 The green vertical line denotes the immunity threshold $1/\mR= s_{EEFOA}= s_{EEIn}.$
 Note that    the epidemic will  spend at first a long time (since births and deaths have  slow rates as compared to the disease) in  the vicinity of the manifold $\vi(t) =0$, where the three processes are indistinguishable, before turning towards the  endemic  equilibrium point(s).}
 \label{f:PIP}
\end{figure}

 Fig.~\ref{f:PIP} compares the qualitative behavior and equilibrium points of the $(\s,\i)$ coordinates of the three variants
of a SIR-type example  (discussed  in detail in \cite{AABH}).


We restrict our study to the biological feasible region $$\mD_4=\left\{(\s,\e,\i,\r)\in \mathbb{R}_+^4, \s+\e+\i+\r=1\right\}.$$
Its invariance is ensured by the fact that
\be{inv} \pr{\s+\e+\i+\r}'=( {\de \i -\L-\g_r})\pr{\s+\e+\i+\r-1},\ee
hence $\s_0+\e_0+\i_0+\r_0=1$ implies $\s+\e+\i+\r=1$, for all $t>0$.

We will work in three dimensions by  eliminating $\r=1-\s-\e -\i$.  The first equation of \eqr{SEIRsc} changes then, and the system becomes \be{SEIR3} \bc
\s'(t)= \L+ \g_r - \g_r (\e(t)+ \i(t))  -\s(t) \pp{\Lgg + (\b_i-\de) \i(t)+\beta_e \e(t)}, \\
\bep
\e'(t)\\
\i'(t)\eep =
\pp{\bep  \beta_e \s(t)  -v_1  &  {\b_i\s(t)} \\
 \ei&-v_2\eep+ \de \i(t) I_2}
  \bep \e(t)\\\i(t)\eep
 \\ (\s,\e,\i) \in \mD_3:=\{(\s,\e,\i)\in \R_+^3, \s+\e+\i \le 1\}\ec.
\ee

Concerning fixed points, we note first, following \cite[(4.2)]{LiGraef}, \cite[(13)]{SunHsieh}, the following necessary condition at a fixed endemic point:
\be{L42}
\begin{split}
&  s'+e'+i'=0\Eq e(\g_e-e_i)+\g i+s \g_s=\g_r r+\L r-\de i r\\
& \Eq (1-\pr{e+i+s}) (\Lambda+ \g_r - \de i )=e \left(\g_e-\ei\right)+\gamma  i+s \gamma s \Lra i < i_c:=\min[1,(\Lambda+ \g_r)/\de],
\end{split}
\ee
where the last implications follows since the right hand side of the last equation in \eqr{L42} is positive, due to $ e_r=\g_e-e_i \ge 0$.

Thus, the search for fixed points may be reduced to the domain
\be{De} \mD_f:=\mD_3 \cap \{\i: \i < i_c=\min[1,(\Lambda+ \g_r)/\de]\}.\ee

As a quick preview of the  \ngm\  ``factorization" idea, we note now that
 the disease free  equation
$ \s'= \L  -\s  (\L +\g_s) + \g_r (1-\s)$ (defined by $e=i=0$, $r=1-s$)  has fixed point
\be{sd}\sd=\fr{\L +\g_r}{\Lgg} \Lra \rd=1-\sd=\fr{\g_s}{\Lgg},\ee
whenever $\Lgg\neq0$, which will be assumed \fno. 
Note this formula contains only three of the  parameters.

The other $6$
parameters
 $\g$, $\de$, $\b_i$, $\beta_e$, $\ei$, $\g_e$ intervene in the ``infection equations"  for $\e,\i$, and will determine the basic reproduction number  \eqr{R0}
$$\m0=\fr{\sd}{\Lg} \pp{ {\beta_e  } +
 \frac{ \b_i \ei}{\Lambda +\gamma +\de }}:={\sd}\mR,$$
 where $\mR$ is precisely the $\m0$ which would be obtained in the absence of vaccination.

When $\de=0$, our model \eqr{SEIR3} reduces to the eight parameters constant population model studied in \cite{Ott}, which has a unique endemic point, with coordinates expressible in terms of $\m0$; \fe $\se =\fr{\sd}{\m0}:=\fr 1{\mR}.$ Its global stability is however hard to prove
\cite[Conjecture 15]{Ott}, and the authors resolve only the case $\beta_e <\g$.


\subsection{Some  history of the SEIRS and SAIRS varying population models }\label{s:hys}

A first important paper on the varying population  SEIRS model (recall that both notations SEIRS and SAIRS have been used already for the same model) is  \cite{Green97}, where  $\beta_e=e_r=0$ 
and where
a proportion $q$ of vaccines is allocated to the  new born. We took for simplicity $q=0$, since this parameter does not modify in an essential way the mathematics involved. Note that $\de$, $\L$, $\g_e$, $\g_s$ and $\g_r$ are denoted in \cite{Green97} by $\al$, $r$, $\sigma$, $p$ and $\ep$, \resp.

Besides the typical weak $\m0$ alternative, \cite[Thm 2.3(i)]{Green97} establishes also {\bf global asymptotic stability} (GAS) of \DFE\ when $\m0<1$ and when either
(I) $\fr{\b_i\g_e}{\L+\g_e}\leq \de $, (II) $\g_s=0$,  or a certain non-explicit condition holds.
A second result \cite[Thm 2.3(ii)]{Green97} establishes the uniqueness of the endemic point when $\m0>1, \text{  and }
 \g_r \leq \min[\g_e,\g].$

Stimulated by the many open cases left above, \cite{LiGraef} considered the particular case $\beta_e=e_r=\g_r=\g_s=0$.
 As the authors explain, the difficulties  met, especially in the global stability problem,  forced them to devise  new ingenious  methods.
 \eqr{R0} becomes now
$\m0=
 \frac{ \b_i \g_e}{\left(\g_e+\L\right) \left(\Lambda +\gamma +\de \right)}$  (note that $\b$, $\g_e$, $\L$, $\de$ and $\m0$ are denoted in \cite{LiGraef} by $\l$, $\ep$, $b$, $\al$ and $\si$, \resp, and that $\sd=1$).
 Building on previous results in the limiting cases
$\de=0$ for SEIR models with constant population \cite{LiMul} and   $\g_e\rightarrow\I$  for SIR models with varying population \cite{busenberg1990analysis}, they establish the uniqueness and global  stability of the endemic point
 \cite[Cor. 6.2, Thm6.5]{LiGraef} when $\m0>1$.

 Note that the Corollary is under the assumption of
non-existence of non-constant periodic solutions, and the Theorem {requires the additional assumption}
$\de <\g_e$;  thus  the complementary case was left open.

Next, \cite{LiMul0} studied the case of waning immunity, leaving again  open many cases.
Note that the so-called ``geometric approach to stability" method,  initiated   in \cite{LiGraef,LiMulgeo,LiMul0},  was used many times afterwards -- see \fe\ \cite{LuLu}.

 Eleven years later, \cite{SunHsieh} re-attacked the  \cite{Green97} problem with $\beta_e=e_r=\g_r=0$ and vaccination $\g_s \geq 0 $ (denoted by $ a\si$).  We note, by  adding the two infection rates in \cite[Fig. 1]{SunHsieh}, which yields
 $\l(1- a \si),$  that the model depends only on $ a\si$, and so introducing two parameters for $\g_s$ unnecessarily complicates the mathematics. \Fr the infection rate $\l(1- a \si)$ might as well be denoted by one parameter, and we chose the classical $\b$. Finally,
    one can   relate their results  to ours by substituting $\b$ with $\g$ and setting $\sigma=1$ (i.e. giving up the second source of infections), and  $ \l =\b/(1-\g_s)$.
  Then, 
  \cite[(8)]{SunHsieh}
\be{R}\m0 =\sd\frac{\b_i\gamma _e }{\left(\Lg \right) \left(\Lgn \right)}:=\mR \sd,\ee
where
$$ \mR=\frac{\b_i\gamma _e }{\left(\Lambda+\gamma _e \right) \left(\Lgn \right)} \text { and } \sd=\fr \L{\L + \g_s}.$$
See also \eqr{R0}.
\beR Let us note two open problems left in \cite{SunHsieh}. \BEN \im  \cite[Thm 2.1]{SunHsieh} proves the  global stability of the DFE  when
$$\mR\leq 1$$
(by using the Lyapunov function $ \gamma_e e+ (\gamma_e +\Lambda) i$,
which is  different from the Lyapunov function used in  \cite{Shuai}, $e (\gamma +\Lambda +\de)+\gamma_e i$. 
 {This leaves  open the  case  $\m0 <1 <  \mR$, when the DFE is locally, but perhaps not globally stable, and also the question of choosing the ``most performant" Lyapunov function.}

\im \cite[Thm 2.1]{SunHsieh} establishes global stability of a unique endemic point, in certain cases, and suggest
that outside those cases, ``there may exist stable periodic solutions".
\EEN

\eeR

 Seven years later, \cite{Brit18} revisited the \cite{LiMul0} problem with $\g_s = 0 $  and waning immunity $\g_r\geq 0$ (denoted by $\rho$).
 The authors remark that in this particular case, there are still open questions: ``We have not proved, but we
strongly believe that if $\m0>1$, then the system \eqr{SEIRsc} has one and only one endemic equilibrium, and that this equilibrium is globally asymptotically stable.''
The \cite{LiMul0} problem was revisited then in \cite{LuLu}, who claim  to have removed also the restriction $\de <\g_e$ of \cite{LiGraef};  {however,  their crucial equation \cite[(3.28)]{LuLu} is wrong, See Appendix}.
The fact  that the papers \cite{SunHsieh,LuLu}
contain  unnecessary conditions,  mistakes and wrong conjectures suggests that for complicated epidemic models with infection rates depending on $N$,  ingenuity stops being enough, and it must be accompanied by
 verifications via symbolic software,  rendered public, as is the case with our paper (in line with the so-called ``reproducible research" \cite{leveque2012reproducible}).

Finally, \cite{Ott} apply the geometric approach to global stability, taking into account  vaccination and loss of immunity, and also the possibility that the exposed, or rather the asymptomatic, are infectious, but for a simplified constant population model with $\de=0, \L\ge 0$.  
The basic reproduction number \eqr{R0}
becomes now \cite[(5)]{Ott}
$$\m0=\pp{ {\beta_e  } +
 \frac{ \b_i \ei}{\Lambda +\gamma}}\fr{\sd}{\g_e+\L}$$
 (with the correspondence $\L\rightarrow\mu$, $\beta_e\rightarrow\beta_A$, $\ei\rightarrow\al$, $e_r\rightarrow\de_{A}$, $\g \rightarrow\de_{I}$, $\g_r\rightarrow\g$, $\g_s\rightarrow\de$).
 The global stability is proven only in the case $\beta_e<\g$.

These multiple but only partially  successful attempts at solving the \cite{Green97} problem    motivated us to revisit it.\fn[5]{Yet another motivation was the literature on bifurcation analysis  of SIR and SEIR-type models \cite{Liu1986,Wang,Ruan,Vyska,ZhouFan,Gupta}, and \cite{Liu87,douris2019global}   which contains yet another plethora of open problems}

\subsection{Warm-up: the weak $\m0$ alternative for the  DFE equilibrium   \label{s:R0} }
Note that the DFE \eqr{sd} is a stable point for the disease-free equation
$$\s'(t)= \L + \g_r  -\s(t) \pr{\Lgg}. $$
Then, \wmc the quadratic - linear decomposition of $$
 \bep\e'(t)\\\i'(t)\eep =
\bep  \beta_e \s(t) + \de \i(t) -(\g_e +\L)  &  {\b_i\s(t)} \\
 \ei& \de \i(t) -v_2\eep\bep \e(t)\\\i(t)\eep =\mF(\e,\i)-
\mV(\e,\i), $$
where
    \bea
\mF(\e,\i)= \bep  \beta_e \s \e +\b_i\s \i+ \de \e \i\\\de \i^2 \eep , \;
\mV(\e,\i)= \bep (\Lg) \e  \\ -\ei \e   +\i v_2\eep
\eea
satisfies the splitting assumptions of the NGM {(Next Generation Matrix)} method \cite{Van08}.

The partial derivatives at the DFE of $\mF(\e,\i),
\mV(\e,\i)$ are
\be{FV}
F= \sd \bep  \beta_e  & \b_i \\ 0 &  0
\eep,
V= \bep  v_1 & 0 \\ -\ei & v_2   \eep,
\ee
and
 \bea F V^{-1}=
 \bep  \beta_e \sd& \b_i\sd\\ 0 &  0
\eep  \bep  \fr 1{v_1} & 0 \\  \fr {\ei}{v_1 v_2}&  \fr 1{v_2}   \eep=\left(
\begin{array}{cc}\sd\pp{ \frac{\beta_e  }{v_1} +
 \frac{ \b_i \ei}{v_1 v_2}}
 & \frac{\sd\b_i}{v_2 } \\
 0 & 0\end{array}
\right).
\eea
We may conclude then by the \wk\ \ngm\ result of \cite{Van08} that:

\beP \label{walt}
The SAIR/V+S basic reproductive number is
\be{R0}\m0
:=\sd\fr{\beta_t}{v_1 v_2},\ee
where $v_i$ were defined in \eqr{h} and where \be{bt} \b_t:=  \b_i \ei + {\beta_e  }{v_2 }.\ee
\Fr the weak $\m0$ alternative\fn[3]{The strong $\m0$ alternative \cite{Shuai} covers also the case $\m0=1$.} holds, i.e.
 \BEN \im If $ \m0<1 \Eq (\L+\g_r) \beta_t < v_0 v_1 v_2$ then the disease-free equilibrium is locally asymptotically stable.
\im If $ \m0>1 \Eq (\L+\g_r) \beta_t > v_0 v_1 v_2$, the disease-free equilibrium is unstable.
\EEN
\eeP

\beR \la{r:AB}   The formula for $\m0$   follows also from the \cite{Arino} formula for  SIR-PH models with new infection matrix  of rank one $F= s  \ba \vb$ -- see \cite{AAK,AABBGH}, and  below, for the definition of this class. This formula becomes, after including demography \para s $\L,\vn$,
$$\m0=\vb \; V^{-1} \ba,\quad  V=-A+Diag(\vn+ \L \vec 1),$$
  In the particular SAIRS case, this formula may be applied with the parameters defining the model, which are    $$\vb=\bep \beta_e,&\b_i\eep, \ba =\bep 1\\0\eep,A=\bep -\g_e&0\\\ei&-\g \eep,   \vn=\bep 0,&\de\eep.
 $$

 Finally, we will write
 $$ {\vec a}:= \vec  1(-A) =\bep \g_e -\ei,& \g \eep \text{ in this example},$$
 for a quantity which will appear often below. -- See for example \eqr{sairsp}
 {where we used the standard notation in the theory of phase-type distributions.}

 \eeR

 A direct proof of Proposition \eqr{walt} is also  easy here. Indeed, after reducing \eqr{SEIRsc}   to a three order system by $r=1-s-e-i$,  the  Jacobian matrix is
\bea
J=\left(
\begin{array}{ccc}
 -\vb \bi + \de i  -\Lambda -\gamma _r-\gamma _s & -s \beta_e-\gamma _r & s (\de -\b_i)-\gamma _r \\
 \vb \bi  & s \beta_e+i \de -v_1  & e \de +\b_i  s \\
 0 & \ei & -\gamma +(2 i-1) \de -\Lambda  \\
\end{array}
\right),
\eea
(where $\vb \bi =\beta_e e +\b_i i$),
and
\small
\bea
J_{DFE}= \bep
 -v_0 & -\beta_e \sd-\gamma _r & (\de -\b_i) \sd-\gamma _r \\
 0 & \beta_e \sd-v_1  &\b_i\sd \\
 0 & \ei & -v_2  \\
\eep.
\eea
Note  the evident block structure $\{s\}, \{e,i\}$ (which is the driving idea behind the \ngm\ method),
with one  negative eigenvalue $-v_0$.

Now  the ``infectious" determinant    $v_1 v_2 -  \sd {\beta_t } $  is positive iff
$$v_1 v_2 (1-\sd \fr{\beta_t }{v_1 v_2}) =v_1 v_2(1-\m0)>0 \Eq \m0<1.$$
This is also the stability condition, since it may be shown that $\sd {\beta_t }<v_1 v_2$ implies also the trace condition $\beta_e \sd-v_1   -v_2 <0$, and so  the $\m0$ alternative holds.

\beR \la{r:cr}  As usual, it is useful to introduce  critical vaccination and critical ``total contact"  (recall $\beta_t={\b_i \ei}+\beta_e v_2$) parameters, as the unique solutions of $\m0=1$ with respect to $\g_s$ and $\beta_t$.
  The critical values are
\be{crv} \g_s^* =(\L+\g_r) \pp{\fr{\beta_t}{v_1 v_2}-1}:=(\L+\g_r) \pp{\mR-1}, \quad \beta_t^*=\frac{v_0 v_1 v_2}{\Lambda +\gamma _r}=\frac{v_1 v_2}{\sd}.\ee

These are particular cases of the SIR-PH formulas \eqr{crvPH}.

\eeR

\subsection{The endemic point for the \FA approximation, and the determinant formula \la{s:LSFA}}
We consider now
the following ``first  approximation" ({\bf FA}) of the {SIR-PH-FA} varying population dynamics

\be{SYRFSA}
\bc
  \bi  '(t)=    \pp{\s(t)  \; B - V} \bi  (t)  \\
  \s'(t)= \L - \s(t)\; \vb \bi (t)  -\pr{ \L + \g_s } \s(t) + \g_r \r(t)    \\
\r'(t) =  \va \bi  (t)+  \s(t) \g_s    -
\r(t)  \pr{\g_r  +\L } \\
(\s(t), \bi (t), \r(t)) \in \R_+^{4}
\ec,
\ee
where $\bi(t):=\bep\e(t)\\\i(t)\eep,  \vb \; \mbox{and}\; \va$ are defined in Remarks \ref{r:AB},  and $V$ in \eqr{FV}.
Here the extra deaths $\de$ due to the disease in state $i$ are kept, but the quadratic interaction terms involving $\de i$ were neglected. Under this approximation, the endemic point and the determinant of the Jacobian have  { elegant formulas in terms  of $\m0$}, some already discovered and others hidden in the particular cases of \cite{RobSti,Ott}.

\beR  For this approximation, the sum of the variables is  constant only if $\de=0$; therefore,    $\r$ may not be eliminated, and we must work in four dimensions.\eeR

\beL a) For the SAIR model \eqr{SYRFSA} with extra deaths $\de$, put $\bi_{ee} =\bep \eE\\\ie \eep$ and
$$\bc v_0:=\L+\g_r+\g_s,\\ v_1:=\L+\g_e,\\ v_2:= \L+\g+\de,\\ v_3 := (v_1+ \g_r) {v_2}  +  \g_r \ei. \ec $$ Then, the following formulas hold at the endemic point:
\be{endSA}
\bc
    \se =& \dfrac{\sd}{\mathcal{R}_0}=\fr 1{\mR},\\
    \bep \eE \\\ie\eep=&  
    {\fr{\L +\g_r}{\L v_3 +\g_r \de e_i}}(1-\frac{1}{\m0}) \bep  v_2\\
      \ei  \eep, \\
   \vb \bi_{ee}=& \b_t \fr{\L +\g_r}{\L v_3 +\g_r \de e_i}(1-\frac{1}{\m0})= \frac{\L (\m0-1) v_0 v_1 v_2}{\L v_3 +\g_r \de e_i}=:EI,\\
   \va \bi_{ee}=& \fr{\L +\g_r}{\L v_3 +\g_r \de e_i}(1-\frac{1}{\m0})  \, \pp{\g_e \g + (\L+\de)e_r},
    \\\re=& \Bigg(\g_s\Big(\L+v_2+e_i(\L+\de)\Big)+\L(\mR-1)\pr{v_2 e_r+\g e_i}\Bigg)\times \frac{1}{\mR\pr{\L v_3 +\g_r \de e_i}}.
    \ec
\ee

b) All the coordinates  are positive iff $\m0>1$.

c) The vector $\bi_{ee}$ checks   the general SIR-PH normalization formula \eqr{inorm}.

\eeL

\Prf a) One can do a  direct computation, or apply \cite[Prop. 2]{AABBGH}, a particular case of which is included for completeness as Proposition \ref{p:ee}, Section \ref{s:SIRPH}. That result is expressed in terms
of the \PF eigenvector of the matrix $M=\fr 1{\mR} B -V$ (for the $0$ eigenvalue); in our case, this is $\bep v_2\\\ei \eep$,  see second equation in \eqr{endSA}.

b) Is obvious.

c)   May be easily checked,  combining  the last two rows in \eqr{endSA}.

\QED

 \beR
 Several particular cases of this problem have been studied in the literature. The case $\de=0$ is studied in \cite{Ott}, and
 \cite{sun2012}
considered the case $\b_e=\g_r=e_r=0,\de>0$,
where
 \bea \ba =\bep 1\\0\eep,A=\bep -\g_e&0\\\g_e&-\g \eep, {\vec a}^t:= (-A)^t \bff 1
 =\bep 0\\\g \eep, \vb=\bep 0&\b_i\eep, \vn=\bep 0&\de\eep.
 \eea
The normalization formula \eqr{inorm}  reduces in this case to
 $$ {\vb \bi= \L \pp{\mR - \frac{1}{\sd}}=\Lambda  \left(\frac{\b_i \gamma _e}{(\Lgn) \left(\gamma _e+\Lambda \right)}-1\right)-\gamma _s}.$$

\eeR

%% file: SAIR2.tex
\beL \la{l:id} {The following  remarkable identity holds} 

\be{detO} Det(J_{ee})={\L} v_0 v_1 v_2(1-\m0)=-Det(J_{dfe}), \ee
where $v_i$ are defined in \eqr{h}.

    \eeL

    \prf  See the proof of Proposition \ref{p:ee}.2.
    
      \beR  We conjecture that the endemic point is always locally stable  when $1<\m0 $.
We have attempted to apply  the classic Routh-Hurwitz-Lienard-Chipart-Schur-Cohn-Jury (RH) methods \cite{Routh,Jury,Daud}, which are formulated in terms of the coefficients of the characteristic polynomial $Det(J-z I_n)=(-z)^n+ a_1 z^{n-1}+...+ a_n$, and of certain Hurwitz determinants $H_i$ \cite[(15.22)]{Routh}.
At order four, $Det(J-z I_3)=z^4- Tr(J) z^{3}+z^2 M_2(J) - z M_3(J)+  Det(J)$,
where $M_2,M_3$ are the sums of the second and third order principal leading minors of $J$, and one ends up with \cite[pg. 137]{Routh}
    $$\bc Tr(J)<0, M_2 >0, M_3 <0, Det(J)>0,\\
0 <Tr(J) \pr{ M_2 M_3-Tr(J) Det(J)}-M_3^2
\ec.
$$

 Now in our example  the determinant is positive when $\m0>1$ by Lemma \ref{l:id}, and the trace, given by {\bea
    -\frac{\g_r e_i(\g_s+\L)(\L+\de)+\L v_2\pr{v_1(\mR(\L+\g_r)+\L)+\g_r(\L+\g_s)}}{e_i \g_r(\L+\de)+\L v_2(\g_r+v_1)}-v_2- (\L+\g_r)-\frac{e_i \b_i}{\mR v_2}
\eea     is negative.

\How the check of the sum of the second and third order principal leading minors of the Jacobian at EE, and of the additional  Hurwitz criterion,  seemed to exceed our machine power.\fn[4]{At order three, RH becomes $\bc Tr(J)<0,\\
     Tr(J) M(J)<Det(J)<0,\ec$
where $M$ is the sum of the second-order principal leading minors of $J$, which is considerably simpler.}}
\eeR

 \QED



%% file: SIRPH.tex
\section{A review of Arino and rank-one SIR-PH models \la{s:SIRPH}}

\subsection{SIR-PH models with  demography, loss of immunity, vaccination and
one susceptible  and one removed classes \label{s:SIRPHdef}}

The fundamental concept of \brn\ $\m0$  can  be only defined (as the spectral radius of the \ngm)  for epidemic models to which the \ngm\ assumptions apply.   It seems  more practical therefore to restrict to ``Arino models" where $\m0$ may be
explicitly expressed in terms of the matrices that define
the  model  \cite{ma2006generality,Arino,Feng,Andr}.

The idea behind these models is to further divide  the noninfected \com s   into
(Susceptible) (or input) classes, defined by producing ``new non-linear infections", and  output R classes (like $D,D_e$ in our first example), which are fully determined by the rest, and may therefore be omitted from the dynamics.
 \Fr it is convenient to   restrict to   epidemic models with linear force of  infection, since it is known that non-linear forces of infection may lead to very complex dynamics \cite{Liu1986,Liu87,georgescu2007global,Ruan,Ruan12,Gupta}, which are not always easy
to interpret epidemiologically.  This is in contrast with the Arino models,
  where typically
  one may  establish the absence of periodic solutions (closed orbits, homoclinic loops and oriented phase polygons) \cite{Raz,Arinovar}.

 It is convenient to   restrict even further  to the case of one removed class (\wlo) and  {\bf only one susceptible class} (a significant simplification).

\beD  A  ``SIR-PH epidemic" {of type $n$}, with demography parameters $(\L,\mu)$ (scalars), loss of immunity and vaccination \paras\ $ \g_r, \g_s$,   is  characterized  by two  matrices $A,B$ of dimensions $n \times n$ and  a column  vector of extra death rates $ \vn$. This model  contains  one  {  susceptible class} $S$, one removed state $R$ (healthy, vaccinated, etc), and a   $n$-dimensional vector of ``disease" states $\bff I $ (which may contain latent/exposed, infective, asymptomatic, etc).   {The dynamics are}:

\begin{align}
S'(t)&=\L N  -\fr{S(t)}N \vb  \bff I (t)   -( \g_s+ \mu) S(t) +  R(t) \g_r, \quad\quad \vb= \vec 1 B ,\nonumber\\ 
\bff I'(t)&=   \pp{\fr{S(t)}N   B + A   -Diag(\vn+ \mu \vec 1)} \bff I (t),\la{SYRN}
\\ R'(t)&= \vec a \bff  I(t)  + \g_s S(t) - R(t) (\g_r +\mu), \;  \vec a=\vec 1 (-A) \nonumber
\\N'(t)&= S'(t)+ \vec 1 \bff  I'(t)  + R'(t) = (\L -\mu) N-
\bff I(t) \vn, \nonumber
\\D'(t)&=   \mu(  S(t) + \vec 1 \bff I(t) + R(t)) ,\nonumber
\\D_e'(t)&=   \bff I(t) \vn  .\nonumber
\end{align}

\QED
\eeD

Here, 
\BEN
\im $\bff I(t) \in \mathbb{R}^n$ is a row vector whose components model a set of
disease states (or classes).

\im $R(t) $  accounts for individuals who recovered from the
infection.

\im $ B $ is a $n \times n$ matrix, where each  {entry} $B_{i,j}$ represents the
force of infection of the  {disease} class $i$ onto class $j$.
 {We will denote by $\vb$ the vector containing the sum of the entries in each row
of $B$, namely, $\vb=\vec 1 B $.}

\im  $A$ is a $n\times n$ Markovian sub-generator matrix (i.e., a Markovian generator
matrix for which the sum of at least one row is strictly negative),  {where each
off-diagonal entry $A_{i,j}$, $i\neq j$, satisfies $A_{i,j}\geq 0$ and describes the
rate of transition from disease class $i$ to disease class $j$; while each diagonal
entry $A_{i,i}$ satisfies $A_{i,i} \leq 0$ and describes the rate at which individuals
in the disease class $i$ leave towards non-infectious compartments.}
Alternatively, $-A$ is a  non-singular M-matrix \cite{Arino, Riano}.\fn[4]{An M-matrix
is  a real matrix $V$ with  $ v_{ij} \leq 0, \forall i \neq j,$ and having eigenvalues
whose real parts are nonnegative \cite{plemmons1977m}.}

\im $\vn \in \mathbb{R}^n$  is a row vector describing the death rates in the  {disease  \com s, which are  caused by the epidemic.}

\im $\g_r$ is the rate at which individuals lose immunity  (i.e. transition from recovered states to the susceptible state).

\im $\g_s$ is    the rate at which individuals are vaccinated (immunized).

\EEN

\beR \la{e:va}
\BEN \im Note that  $\vec a^t:=(-A)^t \bff 1$ is a vector with a \wk\   probabilistic interpretation in the theory of \PH distributions: it is the column vector which completes a matrix with negative row sums to a matrix with  zero row sums.

\im A  particular but revealing case
is that when the matrix $B$ has rank 1, and is necessarily hence of the form $B=\ba \vb$, where $\ba \ \in \mathbb{R}^n $ is a {\bf probability column vector} whose components $\a_j$ represent the fractions of susceptibles entering into the  disease compartment $j$, when  infection occurs.  We  call this case ``rank one SIR-PH", following Riano \cite{Riano}, who emphasized its probabilistic interpretation -- see also \cite{Hurtado}, and see \cite{hyman1999differential} for an early appearance of such models.

\EEN
\eeR

It is  convenient to reformulate \eqr{SYRN} in terms of the fractions
normalized by the total population
\begin{align}
\label{eq:fractions}
\s&=\fr{S}N, &
\bi &=\fr{1}N \bff I,&
\r&=\fr{1}N R, & N= \s + \vec 1 \bi   + \r.
\end{align}

The reader may check that  the following equations hold for  the scaled variables:
\be{SYRsc}
\bc
 \s'(t)= \L -\pr{ \L + \g_s } s(t) + r(t) \g_r- s(t)\; \pr{ \vb -  \vn}  \bi (t) \\
  \bi  '(t)=     \pp{s(t)  \; B+ A- {Diag\pr{\vn+\L \vec 1} +\vn \bi (t)  I_n}
 }\bi  (t) \\
\r'(t) =   s(t) \g_s +  \vec a  \bi  (t) - r(t)  \pr{\g_r  +\L} + r(t) \vn \bi (t)  \\
s(t) +  \vec 1 \bi (t) + r(t) =1
\ec,
\ee
and the Jacobian, using $\nabla \vn \bi (t)  \bi (t)=  \bi (t) \vn +\vn \bi (t)  I_n$,  is
{\small
\be{Jac} J=
\bep -\vb \bi -(\L+\g_s) &-\s\pr{\vb-\vn}   &  \g_r\\
 B \bi  & \s B  -V+  \bi \vn  &0\\
\g_s   & \vec a +  \r \vn & { -(\g_r  +\L)  }\eep +  \vn \bi I_{n+2}.
\ee
}

By letting $\n:=\s+\vec 1\bi  +\r,$ we have
$$\n'(t)=(\L -\vn \bi (t) )(1-\n(t))=0;$$
The above equation guarantees that  if
$\s(t_0)+\i(t_0)+\r(t_0)=1$ for some $t_0 \in \realnneg$, then
$s(t)+i(t)+r(t)=1$ for all $t \geq t_0$.
Accordingly, in what follows we will always assume that $n(t_0) =1$,
which guarantees that $\n(t)=1, \for t$.

The following definition puts in a common framework the dynamics for the scaled
process and two interesting approximations.

\beD \la{d:fisg} Let $\Phi_s, \Phi_i, \Phi_r \in \{0,1\}$ and let
\begin{align}
\label{SYRsc-def}
 \s'(t) &= \L -\pr{ \L + \g_s } s(t) + r(t) \g_r
- s(t) \bi (t) \vb + \Phi_s s(t)  \vn \bi (t) ,
\nonumber\\
\bi  '(t) &=  \pp{s(t)  \; B
+ A
- Diag\pr{\vn+\L \vec 1}
\Big)
+ \Phi_i   \vn \bi (t)} \bi  (t) ,
\nonumber\\
\r'(t) &=   s(t) \g_s +   \vec a \bi  (t) - r(t)  {Diag \pr{\g_r  +\L \vec 1}  }
+ \Phi_r r(t) \vn \bi (t) ,
\nonumber\\
s(t) &+ \vec 1 \bi (t)  + r(t) =1.
\end{align}

\BEN
\im The model \eqref{SYRsc-def} with $\Phi_s=\Phi_i=\Phi_r=1$  will
be  called scaled model (\SM).

\im  The model \eqref{SYRsc-def} with $\Phi_s=\Phi_i=\Phi_r=0$  will
be called first approximation (\FA).

\im  The model \eqref{SYRsc-def} with $\Phi_s=\Phi_r=1$, $\Phi_i=0$
will  be called  intermediate approximation (\IA).
\QED
\EEN
\eeD

\beXa The classic SEIRS  model
\be{seirs} \bc
\s'(t)= \L   -s(t) \pr{\b_ii(t)+\g_s+{\L}} + \g_r r(t)\\
\bep \e'(t)\\\i'(t)\eep =
\bep  {-(\g_e +\L)} &  {\b_is(t)} \\
 \g_e& -\pr{\g   +\L+\de}\eep
  \bep \e (t)\\i(t)\eep
\\
\r'(t)=  \g_s s(t)+ \g   i(t)- r(t)(\g_r +\L) \ec .
\ee
 is a particular case of SIR-PH-FA model
 obtained when $$\bff \a =\bep 1\\0\eep, A=\bep -\g_e&0\\\g_e&-\g \eep, \vec a=\vec 1 (-A)
 =\bep 0 &\g \eep, \vb=\bep 0 & \b_i\eep, \; \mbox{so}\; B=\bep 0 & \b_i\\ 0 & 0 \eep , \vn=\bep 0&\de\eep.$$
 The SAIR is obtained by modifying the parameters to
 \be{sairsp}\ba =\bep 1\\0\eep, A=\bep -\g_e&0\\ \ei&-\g \eep, \vec a=\vec 1 (-A)
 =\bep e_r &\g \eep, \vb=\bep \beta_e & \b_i\eep, \; \mbox{so}\; B=\bep \beta_e & \b_i\\ 0 & 0 \eep , \vn=\bep 0&\de\eep.\ee

\eeXa

\subsection{The eigenstructure of the Jacobian for the SIR-PH scaled model
}
For the scaled model, we can    eliminate $\r$. Then, the system becomes then $n+1$ dimensional:

\be{SYRsc2}
\bc
 \s'(t)= \L +  \g_r -\pr{ \Lgg } s(t)-\g_r \vec 1 \bi (t)  - s(t)\; \pr{ \vb -  \vn}  \bi (t) \\
  \bi  '(t)=     \pp{s(t)  \; B+ A- {Diag\pr{\vn+\L \vec 1} +\vn \bi (t)  I_n}
 }\bi  (t) \\
s(t) +  \vec 1 \bi (t) \le 1
\ec.
\ee

The Jacobian matrix of the scaled model  is given by

\beq \la{Jac2} \no J&&=
\bep -(\Lgg) -  \pr{\vb- \vn } \bi &-\g_r \vec 1 - \s \pr{\vb- \vn }  \\
 B \bi  & \s B  + A-Diag(\vn+ \L \vec 1)+ \vn \bi  +\vn \bi I_n \eep\\&&:=
\bep -(\Lgg) -  \pr{\vb- \vn } \bi &-\g_r \vec 1 - \s \pr{\vb- \vn }  \\
 B \bi  & \s B  -V + \vn \bi  +\vn \bi I_n\eep,
\eeq
and {
\bea  J_{dfe}=
\bep -v_0 &-\g_r \vec 1 - \s \pr{\vb- \vn }  \\
\bff 0  & \sd B  -V \eep,
\eea
}
where $V:=Diag(\vn +\L \vec 1)-A$.

\beR \la{r:detPH} Note the block structure (which suggested probably the \ngm\ approach), and that
$$Det(J_{dfe})=-v_0 \; Det(\sd B  -V).$$ \eeR

We highlight next a simple but important consequence of the fact that $V=Diag(\vn +\L \vec 1)-A$ is an invertible matrix, especially when $B$ is assumed  to have rank $1$.
\beL \la{l:R1}
a) When $B= \ba \  \vb$ is of rank $1$, the matrix $B V^{-1}$ has precisely one non-zero \eig.

b) The remaining \eig\ equals the trace  $$ Tr(B V^{-1})= \vb  \ V^{-1} \ba :=\mR$$

Hence, the Perron-Frobenius eigenvalue of $B V^{-1}$
is
$$\l_{PF}\pr{B V^{-1}}= \mR.$$
\eeL

\Prf a) Since $B= \ba \vb$ has rank 1,   the same holds for $B V^{-1}$,
and the \textit{"rank-nullity theorem"} $rank(B V^{-1})+ nullity(B V^{-1})= n$
\cite{horn2012matrix} implies that $(n-1)$ of the eigenvalues of $B V^{-1}$ are zero.

b) Using the  invariance of the trace  under cyclic permutations, we conclude that the trace of $  \ba \vb \; V^{-1}$  equals   $ \vb  \ V^{-1} \ba$.
Since $ V^{-1}$ has only nonnegative entries, this value must be positive and hence the Perron-Frobenius eigenvalue.

\QED

\ssec{The  \brn\ for  SIR-PH, via the \ngm\ method \cite{Van,Diek}
\la{s:R0PH}}

We follow up here on a remark preceding \cite[Thm 2.1]{Arino}, and show in the following proposition that their simplified formula for the \brn\ still  holds when loss of immunity and vaccination are allowed, provided that $B=\ba \vb$ has rank one.

\beP \la{p:brn}  Consider a SIR-PH model \eqr{SYRsc},
 with parameters
$$(A, B,\L ,\vn, \g_s, \g_r).$$

\BEN \im
 The unique disease-free equilibrium is $ (s_{dfe},\vz, r_{dfe})=\left(\frac{\L+\g_r }{\Lgg}, \vz,   \frac{\g_s }{\Lgg}\rt).
 $

\im   The DFE is locally asymptotically stable if $\m0< 1$ and is unstable if $\m0 > 1$, where
\be{stab} \m0= \l_{PF}(F V^{-1}),\ee
 where $F=\sd B$, $V=Diag(\vn+ \L \vec 1)-A$ (see \eqr{Jac2}),  and $\l_{PF}$ denotes the (dominant) \PF\ eigenvalue.

\im  For $B:= \ba \vb $  of rank one, we further have
\BEN \im
\be{nRV}\m0= s_{dfe}\; \mR, \text{ where } \mR=   \ba \ V^{-1} \; \vb.   \ee

\im The critical vaccination   defined by solving $\mathcal{R}_0=1$ with respect to $\g_s$ is given by
\be{crvPH}  \g_s^*:= (\L+\g_r)\pr{\ba  V^{-1} \bff b -1}= (\L+\g_r)\pr{\mR -1}. \ee

\EEN

 \EEN
\eeP

\Prf 1. The disease free system ( {with $\i=0,\r=1-\s$}) reduces to
\be{SIRDF}
\s'(t)= \L   -(\g_s+\L) s(t) +( \g_r  s(t)) (1-s(t)).
\ee

2.  It is enough to show  that the conditions of \cite[Thm 2]{Van08} hold.

The DFE and its local stability for the disease-free system have already been checked in the SAIR/V+S example.

We provide now a splitting for the infectious equations:
\bea \bi '(t) &=& \pp{s(t)  B + \vn \bi (t)  I_n } \bi (t)
-\pp{Diag(\vn+ \L \vec 1)-A }\bi (t):=\mathcal{F}(s,\bi )-
\mathcal{V}(\bi ) \eea
 (where $\r=1-\s- \bi  \bff 1$).
 The  \corr\ gradients at the DFE $\bi =0$ are
\be{FVPH} \bc F= \left[\frac{\partial \mathcal{F}(X^{(DFE)})}{\partial \bi }\right]=\s B \\
 V=  \left[\frac{\partial \mathcal{V}(X^{(DFE)})}{\partial \bi }\right]= Diag(\vn+\L \vec 1)-A.
 \ec \ee
We note that $F$  has  non-negative \elt s, and that $V$ is a  M-matrix, and therefore $V^{-1}$  exists and has  non-negative \elt s, $\for \L, \vn$.
We may check that the \ngm\ conditions \cite{Van08} 
are \satd.

For example,  the  last non-negativity condition 
 \be{par} \bi (t)\pp{Diag(\vn+ \L \vec 1)-A  } \bff 1 \ge 0, \for \bi  \in \mD, \ee
is  a consequence of $-A$ being a  M-matrix, which implies $-A \bff 1 \ge 0,$ componentwise.

3.a)   Using Lemma \ref{l:R1} and the obvious linearity in $\s$, we may conclude that
 $$\m0= \l_{PF}(F V^{-1})= \sd \l_{PF}(B V^{-1})=\sd \mR.$$

3.b). May be  easily verified

\QED

%% file: SIR2.tex
\ssec{The endemic point of the   SIR-PH-FA model \label{s:EqP}}

In this section, we give more explicit results for the 
endemic
equilibrium of the following approximate model, referred to as SIR-PH-FA
\be{SYRscF}
\bc
  \bi  '(t)=    \pp{\s(t)  B - V} \bi  (t)  \\
  \s'(t)= \L - \s(t)\; \vb \bi (t)  -\pr{ \L + \g_s } \s(t) + \g_r \r     \\
\r'(t) =   \va \bi  (t)+  \s(t) \g_s   -
\r(t)  \pr{\g_r  +\L } \\
(\s(t), \bi (t), \r(t)) \in \R_+^{n+2}
\ec.
\ee

\beR For this approximation, the sum of the variables is  constant only if $\de=0$; therefore,    $\r$ may not be eliminated.
\eeR

If $\m0 >1$, then   \eqr{SYRscF} may have  a second  fixed point within its forward-invariant set. This endemic fixed point  must be such that the quasi-positive  matrix $\se  \; B -V$ is singular, and that
 $\bi $ is a Perron-Frobenius positive eigenvector.

  Let $\bi_0$ denote an arbitrary positive solution of
\be{iPF}   (\se \; B  -V)\bi _{0}:=  M \bi _{0}=0,  \ee
and let \be{ii0} \bi=\fr{\L \pp{\frac{1}{\se} - \frac{1}{\sd}}}{\pr{\vb- \frac{1}{\se} \frac{ \g_r}{\g_r +\L} \vec a} \bi_0} \bi_0\ee
denote the  unique vector of disease components $ \bi _{ee}$ which \sats\
 also  the  normalization:
\be{inorm}     \pr{\vb- \frac{1}{\se}  \frac{ \g_r}{\g_r +\L} \vec a} \bi= \L \pp{\frac{1}{\se} - \frac{1}{\sd}}.\ee

\beP \la{p:ee}
Consider a SIR-PH-FA model \eqr{SYRscF}
 with parameters
$(\L,   A, B, \vn,   \g_s, \g_r),$
   where $A$ is assumed irreducible, with $\m0 >1$. Then:

\BEN \im    There exists  a unique endemic fixed point within its forward-invariant set $\R_+^{n+2}$
iff
$$\pr{\vb- \mR \vec a \frac{ \g_r}{\g_r +\L}} \bi_0>0.$$
   This  fixed point   \sats\

\BEN \im  $ \se=\fr 1 {\mR},$  \im  that $\bi $ given by \eqr{ii0} is a Perron-Frobenius positive eigenvector of the quasi-positive singular matrix $\se  \; B -V$,
\im the normalization \eqr{inorm}.
\EEN

\im The determinant identity $Det(J_{dfe})=-Det(J_{ee})$ holds.
\EEN

\eeP

\Prf Recall the fixed point system
\bea  \no
&&\bc   0=       (\se  \; B -V)\bi \\
 0= \L -\pr{ \L + \g_s } \se + \re \g_r- \se\; \vb \bi,    \\
0 =     \va \bi +  \se \g_s - \re (\g_r+ \L ) \Lra \re=\fr 1{\g_r+ \L} \pr{\va \bi +   {\g_s}\se}
\ec\eea
and note that $\re$ is positive if $\bi, \se$ are.

1.  Let us  examine the two cases  which arise from factoring the disease equations. \Mp  we will search separately in the {\bf disease free set} $\{\bi =\vz\}$ and in its complement. Then: A)   either $\bi \in \{\bi =\vz\}$ and  solving
\bea
\bc
 \L = \pr{ \L + \g_s } \s - \r \g_r  \\
\vz =    \s \g_s - \r (\g_r  + \L )
\ec=\bep \s &\r \eep \bep { \L + \g_s } & \g_s  \\
-  \g_r&  - (\g_r  + \L )\eep
\eea
for $\s, \r$ yields the unique DFE,  or

B)  the determinant of the resulting homogeneous linear system  for $\bi  \neq \vz$ must be $0$, which implies that $\s=\se $ \sats\
\be{lP} det\pp{\se   \; B  -V}=0. \ee

Using that $V=Diag(\vn +\L \vec 1)-A$ is an invertible matrix,
and   $det (U U') = det( U)  det( U')$, \eqr{lP} \mbw
\be{ls}
 Det \left[\pr{\se  B - V} V^{-1}\right] = 0 \Eq Det \left[B  V^{-1} -\frac{1}{\se } \mathbf{I}\right] = 0.
\ee
Thus:

a)  $\frac{1}{\se } $ must equal the \PF\ \eig\ $\mR$ (recall Lemma \ref{l:R1}).
Note that $\se  <1$   follows from  $\mR \ge \m0 >1$.

   b)  $\bi $ is a Perron-Frobenius  eigenvector of the quasi-positive matrix $\se  \; B -V$, and hence may be chosen as positive.

c) To determine the proportionality  constant,  it  remains to solve the second equation :
\bea &&
  \pr{ \vb \se-\va \fr{ \g_r} {\g_r +\L }} \bi  = \L -\pr{ \L + \g_s  } \se +    \fr{ \g_r} {\g_r +\L }\g_s \se \Eq  \pr{ \vb -\va \fr{ \mR \g_r} {\g_r +\L }} \bi  = \\&& \L \mR -\L +\g_s \pr{\frac{\g_r}{\g_r +\L} -1}=\L \mR -\L -\g_s \pr{\frac{\L}{\g_r +\L}}= \L \pr{\mR -  \frac{\g_s+\g_r +\L}{\g_r +\L}},
 \eea
 yielding
\eqr{inorm}.


2. At the DFE, the infectious equations decouple,
and the triangular block structure implies  $$Det(J_{dfe})=\L v_0 Det(\sd B -V).$$

For the EE, we will compute the determinant of the  Jacobian matrix:
\bea J=
\bep
  \s B  -V &B \bi  &\bff 0\\
 -\s \vb   &-\vb \bi -(\L+\g_s) &  \g_r\\
 \vec a  &\g_s   & { -(\g_r  +\L)  }\eep, \eea
 after applying   simplifying  row and  column  operations which preserve the determinant (``Neville eliminations" \cite{gasca1992total}), to be denoted by $\propto$.

 But first, we will take  a detour through the more explicit SAIRS-FA model \eqr{SYRFSA}, where $\bi=\bep e\\ i  \eep$, \bea 
J && = \bep
  \b_e s-v_1 &   \b_i s &\b_e e+\b_i i&0 \\
   e_i & -v_2 & 0&0\\
   -\b_e s&-\b_i s &-(\L+\g_s) -\b_i i -\b_e e&\g_r \\
   e_r & \g& \g_s& -(\L+\g_r)
\eep\\
&& \propto \bep
 s \beta_e -v_1  & \b_i s& \b_e e+\b_i i  &0 \\
  \ei &   -v_2 &0&0 \\
   -v_1 & 0 &  - (\L+\g_s)&\g_r\\
   e_r&\g &\g_s&-(\L+\g_r)
\eep
\eea
(here we added row one to row three), and $Det(J_{dfe})$ simplifies to
$$Det(J_{dfe})=\L v_0(v_1 v_2- \sd \b_t)=\L v_0 v_1 v_2(1- \m0).$$

The  Jacobian at the endemic point is
\bea &&J_{ee}= \bep
 \frac{ \beta_e}{\mR} -v_1  & \frac{\b_i}{\mR} & \b_e e_{ee}+\b_i i_{ee}  &0 \\
  \ei &   -v_2 &0&0 \\
   -v_1 & 0 &  - (\L+\g_s)&\g_r\\
   e_r&\g &\g_s&-(\L+\g_r)
\eep \eea
requires more work, and we will start by
 asking Mathematica for the LUDecomposition $J= Per L U \Lra Det(J)=Det(Per) Det(U)$. The first factor $Per$ is a permutation matrix with determinant $-1$ in our case, the second,  $L$,  is lower triangular with one on the diagonal, 
 and the second is upper triangular \bea U=\left(
\begin{array}{cccc}
 e_i & -v_2 & 0 & 0 \\
 0 & \frac{v_2 \g_e-e_i (\Lambda +\de )}{e_i} & \gamma _s & \gamma _s-v_0 \\
 0 & 0 & \frac{\L v_0 v_1 v_2 (1-\m0)}{-\g_r e_i (\Lambda +\de ) +v_2 \Lambda (-\g_r-v_1)} & 0 \\
 0 & 0 & 0 & \frac{-\g_r e_i (\Lambda +\de ) +v_2 \Lambda  (-\g_r-v_1)}{e_i (\Lambda +\de )- \g_e v_2 } \\
\end{array}
\right),
\eea
with determinant
$$-Det(U)=\L v_0 v_1 v_2(\m0-1)=Det(J_{ee}).$$

We provide now a second derivation based on determinant preserving transformations

\bea && J_{ee} \propto \bep
 \frac{ \beta_e}{\mR} -v_1  & \frac{\b_i}{\mR} & \b_e e_{ee}+\b_i i_{ee}  &0 \\
  \ei &   -v_2 &0&0 \\
   -v_1 & 0 &  - v_0&\g_r\\
   e_r&\g &v_0&-(\L+\g_r)
\eep\\&&\propto \bep
 \frac{ \beta_e}{\mR} -v_1  & \frac{\b_i}{\mR} & \b_e e_{ee}+\b_i i_{ee}  &0 \\
  \ei &   -v_2 &0&0 \\
   -v_1 & 0 &  - v_0&\g_r\\
   e_r-v_1&\g &0&-\L
\eep,
\eea
where we  substracted column four from three, and then added row three to row four
and
where $e_{ee}$ and $i_{ee}$ are defined in \eqr{endSA}. We develop now by third column:
\bea &&Det(J_{ee})= EI \bev
  \ei &   -v_2 &0 \\
   -v_1 & 0 &  \g_r\\
   e_r-v_1&\g &-\L
\eev + v_0 \bev
 \frac{ \beta_e}{\mR} -v_1  & \frac{\b_i}{\mR}   &0 \\
  \ei &   -v_2 &0 \\
   e_r-v_1&\g &-\L
\eev \\&&=EI \bev
  \ei &   -v_2 &0 \\
   -v_1 & 0 &  \g_r\\
   e_r-v_1&\g &-\L
\eev - v_0 \L \bev
 \frac{ \beta_e}{\mR} -v_1  & \frac{\b_i}{\mR}   \\
  \ei &   -v_2 \eev\\
  && =EI \bev
  \ei &   -v_2 &0 \\
   -v_1-\g_r & 0 &  \g_r\\
   0&-(\L+\de) &-\L
\eev - v_0 \L \bev
 \frac{ \beta_e}{\mR} -v_1  & \frac{\b_i}{\mR}   \\
  \ei &   -v_2 \eev\\
  && = EI \pr{e_i \g_r(\L+\de)+v_2\L(v_1+\g_r)}+\L v_0\pr{v_2(\frac{\b_e}{\mR}-v_1)-e_i \frac{\b_i}{\mR}}\\ && = EI\pr{\L v_3+ \de \e_i \g_r}
  \eea
where $EI$ is defined in \eqr{endSA}, and the third equality follows by  substracting column three from column one and then adding row one to row three. Then using $\b_e=\frac{\mR v_1 v_2-e_i \b_i}{v_2}$, the last product in the equality four cancels, 
and recall  $EI=\b_e e_{ee}+\b_i i_{ee}=\frac{\L(\m0-1)v_0 v_1 v_2}{\L v_3 +\g_r \de e_i}$, this yields
   \bea Det(J_{ee})=\L v_0 v_1 v_2 (\m0-1)=-Det(J_{dfe}).\eea
  Or,  by developing the first row, the determinant reads
\bea
& Det(J_{ee})= (\frac{\b_e}{\mR}-v_1) \bev  -v_2 & 0 &0\\  0 &  - (\L+\g_s)&\g_r\\ \g &\g_s&-(\L+\g_r) \eev -\frac{\b_i}{\mR} \bev e_i & 0  &0\\  -v_1  &  - (\L+\g_s)&\g_r\\   e_r&\g_s&-(\L+\g_r) \eev\\
& + EI\bev e_i& -v_2& 0\\ -v_1& 0& \g_r\\ e_r& \g&-(\L+\g_r) \eev\\
& =-(\frac{\b_e}{\mR}-v_1)\L v_0 v_2 -\frac{\b_i}{\mR}\L e_i v_0+EI\pr{-\g \g_r e_i+v_2(v_1(\L+\g_r)-e_r\g_r)}\\
& =EI\pr{-\g \g_r e_i+v_2(v_1(\L+\g_r)-e_r\g_r)} \quad (\mbox{by using} \; \b_e=\frac{\mR v_1 v_2-e_i \b_i}{v_2})\\
& =EI\pr{-\g \g_re_i-v_2 \g_r(v_1-\L-e_i)+v_1 v_2(\L+\g_r)}\\
&= EI\pr{v_1 v_2\L+e_i \g_r(v_2-\g)+\L \g_r v_2}\\
&= EI \pr{\L(v_1 v_2+e_i \g_r+\g_r v_2)+e_i \g_r \de }\\
&= EI \pr{\L v_3+e_i \g_r \de},
\eea
thus, $Det(J_{ee})=  \L v_0 v_1 v_2 (\m0-1)=-Det(J_{dfe}).$

\QED


%% file: IA.tex
\section{A glimpse of the intermediate approximation  for the SIR-PH model \la{s:IA}}

The intermediate approximation associated to \eqr{SYRsc} is
\be{SYRIA}
\bc
 \s'(t)= \L -\pr{ \L + \g_s } \s(t) + \r(t) \g_r- \s(t)\;  \pr{ \vb -  \vn}  \bi (t) \\
  \bi  '(t)=    \Big(\s(t)  \; B -V \Big)  \bi  (t) \\
\r'(t) =   \s(t) \g_s +   \va \bi  (t) - \r(t) \Big(\L+\g_r  -\vn \bi (t)) \Big)
\ec;
\ee

\beP \la{p:IA}
\BEN \im
 The DFE points of the scaled, the intermediate approximation, and the FA are equal, given by $(\fr{\L + \g_r}{\L+\g_r+\g_s},\vec 0,\fr{ \g_s}{\L+\g_r+\g_s})$.
 \im An endemic point must \saty\ that $\bi_{ee}$ is a positive eigenvector of the matrix $\se B -V$  for the \eig\ $0$ (same as for the \FA), that
 \bea \bc s_{ee}= \fr 1 \mR, \\ r_{ee}=\frac{\frac{\g_s}{\mR}+\va \bi_{ee}}{\L+\g_r-\vn \bi_{ee}}>0 \ec,  \eea
  and that
 \be{normIA}
 \L(\mR-1)-\g_s +\frac{\g_s \g_r}{\L+\g_r -\vn \bi_{ee}}=\pp{(\vb-\vn)- \mR  \frac{\g_r  }{\L+\g_r -\vn \bi_{ee}} \va} \bi_{ee}.\ee 

Since this equation is quadratic (see \eqr{qno}), we may have a priori two, one or zero endemic points.

 \EEN
\eeP

\Prf \BEN
\im The equations determining the DFE for the three models coincide.
\im  $s_{ee}=\fr 1 \mR$ has been established in the proof of Proposition \ref{p:ee}, and $r_{ee}=\frac{\frac{\g_s}{\mR}+\va \bi_{ee}}{\L+\g_r-\vn \bi_{ee}}$ follows immediately from the last equation in \eqr{SYRIA}. By susbtitution of $r_{ee}$ and $s_{ee}$ into the first equation in \eqr{SYRIA}, we obtain
\bea
&\L(\mR-1)-\g_s +\frac{\g_s \g_r}{\L+\g_r -\vn \bi_{ee}}=\pp{(\vb-\vn)- \mR  \frac{\g_r  }{\L+\g_r -\vn \bi_{ee}} \va} \bi_{ee}, \eea
which yields the result.
\EEN
\QED
\beR
\BEN
\im Under the substitution $\vn \bi_{ee}\rightarrow 0$, \eqr{normIA} reduces to formula \eqr{inorm}.

\im  The existence of positive solutions to \eqr{normIA} requires studying a quadratic equation $X \l^2 + Y \l + Z$. Indeed, putting $\bi_{ee}=\l \bi_{0}$, \eqr{normIA} may also be written as
\beq \la{qno} &\Eq (\L+\g_r -\vn \l \bi_{0})(\L(\mR-1)-\g_s)+\g_s\g_r= (\L+\g_r -\vn \l \bi_{0})(\vb-\vn)\l \bi_{0}-\mR  \g_r\va \l \bi_{0} \no\\
& \Eq \g_s \g_r= (\L+\g_r -\vn \l \bi_{0})\Big[ (\vb-\vn)\l \bi_{0}+\g_s-\L(\mR-1)\Big]-\mR \g_r \va \l \bi_{0},
\eeq
and we find \bea \bc
X=\vn \bi_0 (\vb-\vn) \bi_0,\\
Y=\pr{(\g_s-\L(\mR-1))\vn+\mR \g_r \va+(\L+\g_r)(\vn-\vb)}\bi_0,\\
Z=\g_s \g_r +(\L+\g_r)(\L(\mR-1)-\g_s).\ec\eea 
\EEN
\eeR

\beXa {\bf The intermediate approximation of the SAIRS model} is
\be{sairs} \bc
\bep \e'(t)\\\i'(t)\eep =
\bep  {-(\g_e +\L)+\b_e \s(t)} &  {\b_i\s(t)} \\
 e_i& -\pr{\g   +\L+\de}\eep
  \bep \e (t)\\\i_t\eep
\\
\s'(t)= \L   -\s(t) \pr{\b_i\i_t+\b_e \e(t)+\g_s+\L-\de \i_t} + \g_r \r(t)\\
\r'(t)=  \g_s \s(t)+ e_r \e(t)+\g   \i_t- \r(t)(\g_r +\L-\de \i_t) \ec .
\ee
 is a particular case of SIR-PH-IA model
 ($\ba, A, \vec a, \vb$ and $\vn$ were defined in \eqr{sairsp}).

 For the SAIRS model \eqr{SYRIA} with extra deaths $\de$, putting $\bi_{ee}^{(1,2)} =\bep \eE^{(1,2)}\\\ie^{(1,2)} \eep$, the following formulas hold at the endemic points:
\be{endSI}
\bc
  &  \se = \dfrac{\sd}{\mathcal{R}_0}=\fr 1{\mR}\\
 &  \eE^{(1,2)}= \frac{v_2 e_i \left(v_0 (\de -\mR (\Lambda +\de ))+\Lambda  \de +\Lambda  \mR \left(\Lambda +\gamma _s\right)+\de  \mR \gamma _s\right)+v_2^2 \Lambda  \mR \left(-v_0-v_1+\Lambda +\gamma _s\right)}{2 \de  e_i \left(\de  e_i-v_1 v_2 \mR\right)}\\ & \pm\frac{\sqrt{v_2^2 \left(\left(v_0 e_i (\de -\mR (\Lambda +\de ))+\de  e_i \left(\Lambda +\mR \gamma _s\right)+\Lambda  \mR e_i \left(\Lambda +\gamma _s\right)+v_2 \Lambda  \mR \left(-v_0-v_1+\Lambda +\gamma _s\right)\right){}^2+4 \Lambda  \de  e_i \left(\mR \gamma _s-v_0 (\mR-1)\right) \left(v_1 v_2 \mR-\de  e_i\right)\right)}}{2 \de  e_i \left(\de  e_i-v_1 v_2 \mR\right)} \\
 &  \ie^{(1,2)}=  \frac{v_2 e_i \left(v_0 (\de -\mR (\Lambda +\de ))+\Lambda  \de +\Lambda  \mR \left(\Lambda +\gamma _s\right)+\de  \mR \gamma _s\right)+v_2^2 \Lambda  \mR \left(-v_0-v_1+\Lambda +\gamma _s\right)}{2 v_2 \de  \left(\de  e_i-v_1 v_2 \mR\right)}\\ & \pm\frac{\sqrt{v_2^2 \left(\left(v_0 e_i (\de -\mR (\Lambda +\de ))+\de  e_i \left(\Lambda +\mR \gamma _s\right)+\Lambda  \mR e_i \left(\Lambda +\gamma _s\right)+v_2 \Lambda  \mR \left(-v_0-v_1+\Lambda +\gamma _s\right)\right){}^2+4 \Lambda  \de  e_i \left(\mR \gamma _s-v_0 (\mR-1)\right) \left(v_1 v_2 \mR-\de  e_i\right)\right)}}{2 v_2 \de  \left(\de  e_i-v_1 v_2 \mR\right)}
    \ec.
\ee

 \eeXa

\beR A yet another open problem is whether the endemic point must always exists for the \FA\ and \IA\ models. \eeR

%% file: SAIRsc.tex
\section{Appendix: The scaled SAIRS model: existence, uniqueness, and local stability of the endemic point}\label{s:ee}
\subsection{Reduction to one dimension and the \cite{SunHsieh} problem}
We will follow here the idea of  the particular cases  \cite[(4.3)]{LiGraef}, \cite[(14)]{SunHsieh} and  \cite[(3.28)]{LuLu}, in which the authors eliminate $e,s$ in the fixed point equations
\bea \bc
0= \L+ \g_r(1- e- i)  -s \pp{v_0 + (\b_i-\de) i+\b_e  e}, \\
\bep
0\\
0\eep =
\pp{\bep  \beta_e s  -v_1  &  {\b_i s} \\
 \ei&-v_2\eep+ \de i I_2}
  \bep e\\i\eep\ec
\eea
and study a resulting  polynomial equation in $i$.
{An alternative to the successive eliminations suggested in \cite{LiGraef,SunHsieh,LuLu} is to notice that a strictly positive endemic point must satisfy that the determinant} $$\D=\begin{vmatrix}  \beta_e \s + \de \i -v_1  &  {\b_i\s} \\
 \ei& \de \i -v_2\end{vmatrix}=0.$$
 After eliminating $\s$ from the first equation, the denominator of the determinant is  the denominator of $\s$, and the numerator of the determinant is a fifth order polynomial in $\i$, with  factor $\i$.

\be{sei}e=\frac{i ( v_2-i \de)}{\ei}, \quad
s=\frac{\Lambda +\gamma _r\pr{1-e -i}}
{v_0 +  i(\b_i- \de)+e \beta_e}.\ee
{From the equation of $s$ in \eqr{sei}, we note a typo in \cite[(15)]{SunHsieh}, which should be $s=\frac{\Lambda }
{ i(\b_i- \de) +\L +\g_s}$.}

The next result shows that the existence of endemic points may be  reduced for SAIRS to a one dimensional problem.

\beL An endemic point $E_{ee}$
\satg\ 
$0<\ie<\min[1,(\Lambda+ \g_r)/\de] $ and \be{ce} \g_r(\eE+\ie-1) \leq \Lambda\ee will \saty\ $E_{ee}\in \mD_e$.
\eeL

 \Prf  By \eqr{sei},  $ 0< \ie <1 \Lra \eE>0$. Also, the numerator of $\se$, given by ${\Lambda +(1-\ie-\eE) \gamma _r}$,  is positive by \eqr{ce}.
\Fr $0<\ie< i_c$ {(recall \eqr{L42})} implies that the  denominator of  $\se $ is positive.

 Finally, $ 0 <1-\se-\eE-\ie$ follows from \eqr{L42} and $\se>0$, $\eE>0$, $\ie>0$. \QED

  The next result shows that the endemic point is unique, without the unnecessary extra conditions assumed in \cite{SunHsieh}

  \beP $\m0>1$  ensures the existence and uniqueness of the endemic point
  for the  \cite{SunHsieh}  
 SIR-PH-SM problem.
  \eeP

  \Prf
For the \cite{SunHsieh} problem \eqr{ce} holds trivially (since $\g_r=0$, $\g_s \ge 0$). Thus, it only remains to show
  that the   third order polynomial $p(\i)$ which results from the elimination has precisely one root in  $(0,i_c)$, when $\m0>1$.\fn[1]{The polynomial is of fourth order in general, with a complicated  formula,  
   but the leading coefficient is $-\de^3 \beta_e$, and  when $\beta_e=0$ and $\de=0$,  the fourth order polynomial becomes of third and first order, \resp. 
  }
 It turns out 
  that  this polynomial
 \sats\
$$\bc p(0)=-(\m0-1)\g_e v_1 v_2 \left(\Lambda +\gamma _s\right) <0\\
p(i_c)=\frac{\gamma _e^2 \left(\b_i \gamma  \Lambda +\de  (\gamma +\de ) \gamma _s\right)}{\de }>0\ec.$$
This implies that when $\m0>1$, $p(i)$ must  have either one, two or three roots in  $(0,i_c).$

The last case may be ruled out using an interesting algebraic identity  \cite[(4.3)]{LiGraef}, \cite[(14)]{SunHsieh}
\be{facSH}{p(i)=\m0-f_{SH}(i)}, \quad f_{SH}(i)=f_{Li}(i):=\pr{1+\frac{\b_i- \de}{\g_s+\L} i}\pr{1- \frac{i \de}{v_2}}\pr{1-\frac{i \de}{v_1}}.\ee
This shows that solving the equation $p(i)=0$ (which provides the endemic equilibrium values of $\i$), is equivalent to equating  to $\m0$ a function with known roots $f_{SH}(i).$

As a sanity check,  note  that when $i=0$, this equation reduces to $\m0=1$, which is consistent with the fact  that the endemic point only appears at this threshold (equivalently, when $\m0=1$, the polynomial ${p(i)=\m0-f_{SH}(i)}$ admits only one root, namely $\i=0$).

Refer now to the Figure \ref{f:pltSH} which plots the function  $f_{SH}(i)$. It may be easily checked that the roots of $f_{SH}(i),$ given by $i_1=\frac{v_1}{\de}$,  $i_2=\frac{v_2}{\de}$, $i_3=\frac{\L}{\b_i-\de}$,  do not belong to the range $(0,i_c=\min[1,\frac{\L}{\de}])$ (recall that $\g_r=0$). Indeed, $\i_2>1$, $\i_1>\frac{\L}{\de}$, and   $\m0 >1 \Eq \beta_t > \beta_t^* \Lra \b_i> \de \Lra i_3 <0$.  \Itf the largest root of $f_{SH}(i)=\m0$ must be outside the interval $(0,\frac{\L}{\de})$,  ending the proof. {In \cite[Thm 3.3]{SunHsieh}, the authors used a slightly different approach which leads to several unnecessary cases,  as seen above.} \QED

{From the proof above, when $\m0>1$, the line $f=\m0$ has exactly one intersection $(i_{ee},f_{SH}(i_{ee}))$ with the graph of $f_{SH}(i)$ that satisfies $i_{ee}\in[0,i_c]$,-- see Figure \ref{f:pltSH}.}

 \figu{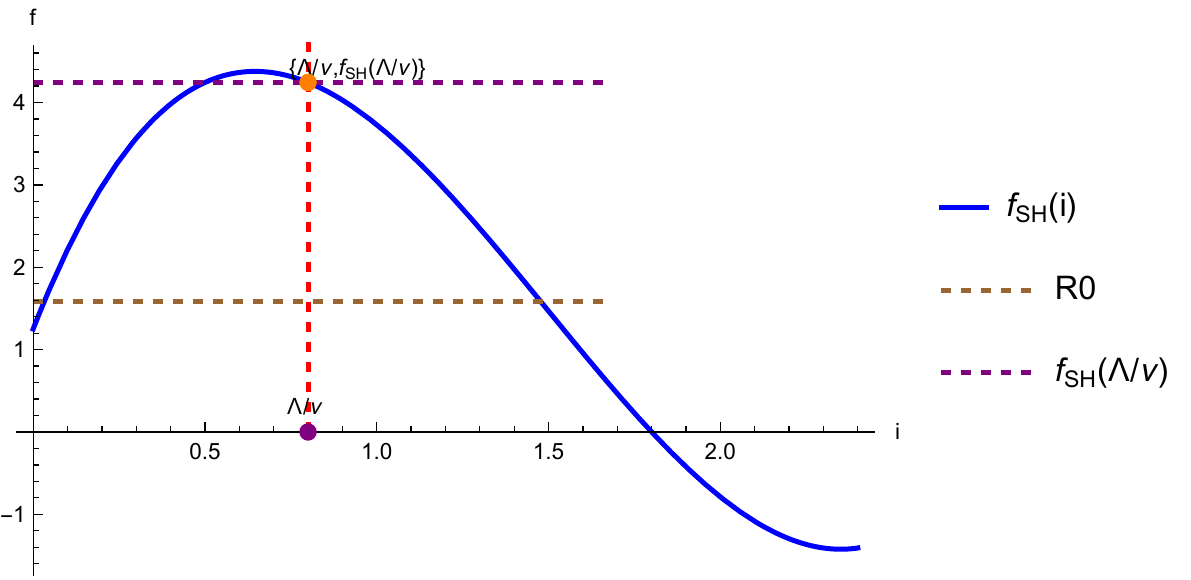}{The existence and uniqueness of endemic point $i_{ee}$ in the interval $[0,i_c]$ for the  \cite{SunHsieh} 
  problem, \cite[Fig. 1]{LiGraef}, 
with $f_{SH}(i)$ as defined in \eqr{facSH}, $f_{SH}(\fr \L \de)= \frac{(\gamma +\de ) \gamma _e \left(\b_i \Lambda +\de  \gamma _s\right)}{\Lambda  \de  (\Lgn) \left(v_1 \right) \left(\Lambda +\gamma _s\right)}$, $\b=18$, $\beta_e=\g_r=0$, $\L=\fr4 5 $ and $\de=\g=\ei=\g_s=\g_e=1$.}{1} 

\begin{figure}[H] 
    \centering
    \begin{subfigure}[a]{0.7\textwidth}
        \includegraphics[width=\textwidth]{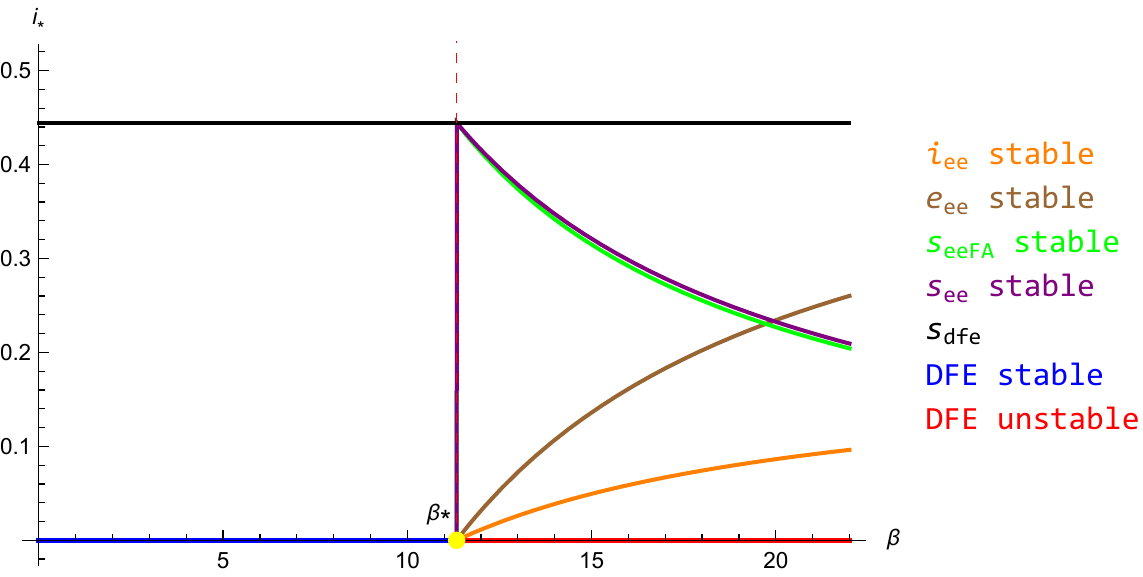}
        \caption{{Forward bifurcation diagram} \wrt\ $\b$, DFE stable from 0 to $\beta_*^{SH}$ where the coordinates of the endemic point are negative and the endemic point is stable after the critical  bifurcation parameter $\beta_*^{SH}$, {with $\beta_*^{SH}:= \frac{(\Lgn) \left(v_1 \right) \left(\Lambda +\gamma _s\right)}{\Lambda  \gamma _e}$, and $\de=1$.}
        \label{fig:bifb}}
    \end{subfigure}
     ~
    \begin{subfigure}[a]{0.7\textwidth}
        \includegraphics[width=\textwidth]{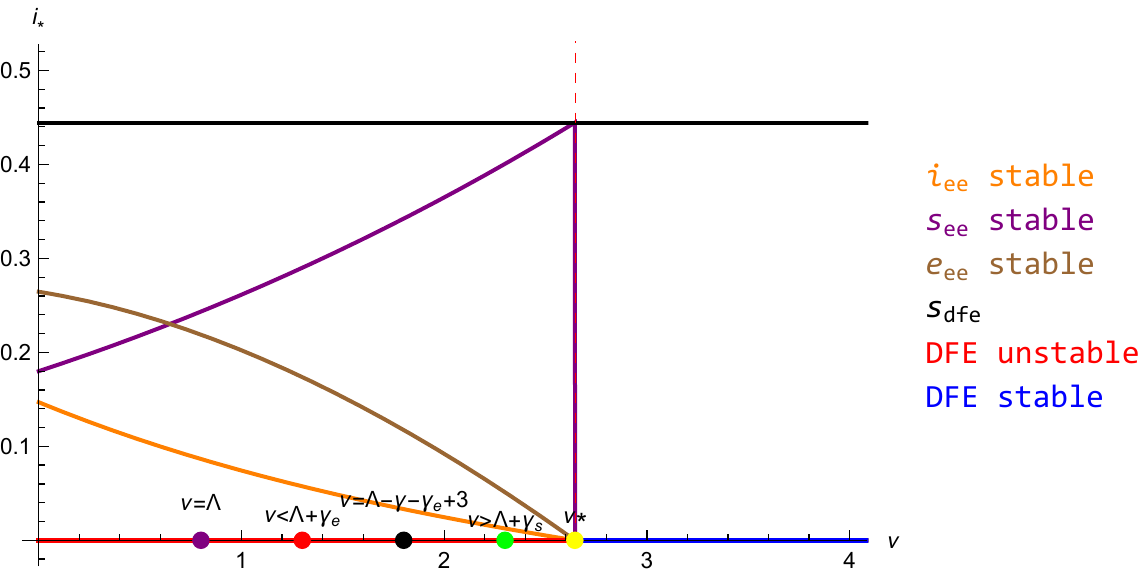}
        \caption{Bifurcation Diagram \wrt\ $\de$, the endemic point is stable  from 0 to $\de_*:=\L \pr{\frac{\b_i \g_e}{(\L+\g_e)(\L+\g_s)}}-\g$ and the DFE is stable when $\de\geq \de_*$, { with $\b=18$. The four points before the yellow one ensure the uniqueness of the endemic point  when $\m0>1$, see \cite[Theorem 3.3]{SunHsieh}.}
        \label{fig:bifnu}}
    \end{subfigure}
    \caption{Bifurcation Diagrams \wrt\ $\b$ and $\de$.
    {Here, $\gamma _r= 0$, $\Lambda = \frac{4}{5}$ and $\gamma =\gamma _e=\ei=\gamma _s= 1$.} %
     \label{fig:bif}}
    \end{figure}

  Figure \ref{fig:bifb} displays the bifurcation diagrams {of the equilibrium value(s) of $\i$} for the \cite{SunHsieh} problem with respect to $\b$;  we note the usual forward bifurcation diagram when $\m0$ reaches its critical value $1$.
Figure \ref{fig:bifnu} displays the bifurcation diagrams {of the equilibrium value(s) of $\i$} with respect to $\de$; we notice that nothing happens in the critical points identified in \cite[Thm 3.3]{SunHsieh}.

    
    Figure \ref{f:PIPSH} compares the qualitative behavior and equilibrium points of the ( s,e+i)-coordinates of  the three variants of the SEIRS scaled model discussed in \cite{SunHsieh}, and note that the endemic coordinates $  s_{EEFA}=  s_{EEIn}=1/\mR,$ which is illustrated by a vertical green line. 

\figu{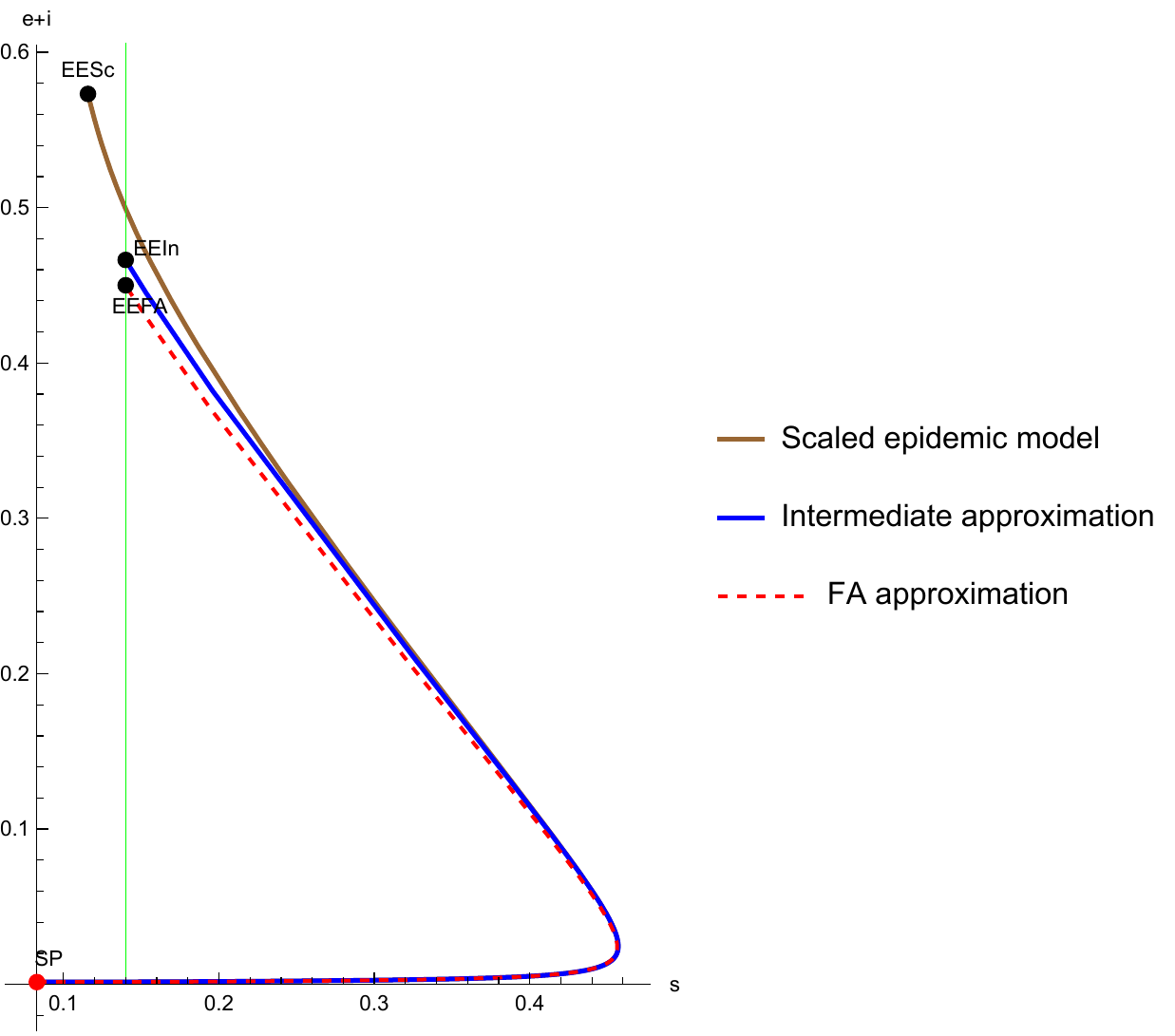}{Parametric Plot of (s,e+i) of the SEIRS scaled model \cite{SunHsieh} and its FA and intermediate approximations  starting from a starting point SP=$(0.0828333,0.0015)$, with $\m0=3.57143,\b=57.1429,\g_r=0,\de=2,\g=\g_e=\g_s=\L=1$. EESc, EEIn, EEFA denote the intermediate points of the scaled model, interdemiate approximation, and FA approximation, respectively. The green vertical line denotes the immunity threshold $1/\mR = s_{EEFA}=  s_{EEIn}$.}{.8}

%% file: con1.tex
\ssec{Conjecture for the SEIRS model}
 The uniqueness of the endemic point and its  local stability  when $\m0>1$ was also  claimed in \cite{LuLu}, and we conjecture that these results are true, even for  SEIRS, with $\b_e=e_r=0, \g_s \ge 0,\g_r\ge 0$.

 However, this must be viewed as an open problem even in the case $\b_e=e_r=0=\g_s$, since  the crucial
  analog of   \eqr{facSH}, the equation \cite[(3.28)]{LuLu} used intensively in their proof is wrong.

  Indeed, recall that the $i$ equation may be written as
  \be{fgLu}
  f_{Lu}(i_{ee})=\m0 +g_{Lu}(i_{ee}), \ee
  where $f_{Lu}(i)=f_{SH}(i)$ (with $\g_s$ replaced by $0$). The equation \cite[(3.30)]{LuLu} states that $g_{Lu}(i)$  is  rational,  which implies that the polynomial $p(i)$  is of fourth order, while the correct order of  $p(i)$ is $3$. In fact,
  \cite[(3.30)]{LuLu}  should be replaced by the  second order polynomial
  \be{gLu}  g_{Lu}(i)= \m0 \frac{\gamma _r-i \gamma _r}{\Lambda }+1 -\frac{\gamma _r (\gamma -i \de +\Lambda +\de ) \left(\gamma _e+i (\b_i-\de )+\Lambda \right)}{\Lambda  (\Lgn) \left(v_1 \right)} \ee

  \figu{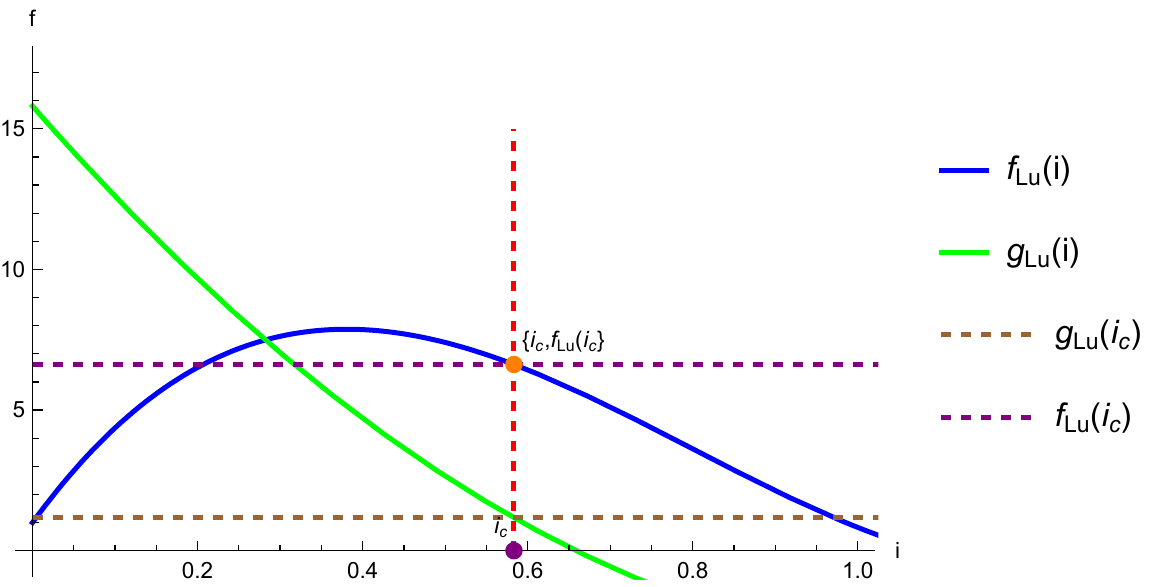}{The existence and uniqueness of the endemic point $i_{ee}$ for the \cite{LuLu} problem, with $i_c=\frac{ 7}{ 12} $ and $\b_i= 9,\Lambda = \frac{1}{6},\de = 2,\gamma = \frac{1}{2},\gamma _e= 2,\gamma _r= 1,\ei= 2$.  
  }{1}

   We conjecture that their result  holds true, since in this case also the third order polynomial
 resulting from the elimination  when $\m0>1$ \sats\
$$\bc p(0)=-(\m0-1)\g_e \Lambda  (\Lgn) \pr{v_1 } <0,\\
p(i_c)=\frac{\b_i \gamma  \gamma _e^2 \Lambda }{\de }>0,\ec$$
and so $p(i)$ must still have either one, two or three roots in  $(0,i_c)$. 

